\documentclass{jfm}
\usepackage{epstopdf, epsfig}
\usepackage{amsmath}
\usepackage{xcolor}
\usepackage{multirow}
\usepackage{longtable}
\usepackage{comment}
\usepackage{hhline}
\usepackage{hyperref}
\hypersetup{
    colorlinks,
    linkcolor={red!50!black},
    citecolor={blue!50!black},
    urlcolor={blue!80!black}
}

\shorttitle{2D cylinders with centre-of-mass offset}
\shortauthor{M.P.A. Assen, J.B. Will,  C.S. Ng, D. Lohse, R. Verzicco, D. Krug}

\title{Rising and settling 2D cylinders with \\
centre-of-mass offset}
 \author{Martin P. A. Assen\aff{1}\aunote{Equally contributing authors},
   Jelle B. Will\aff{1,2}\aunote{} \corresp{\email{jelle.will@unimelb.edu.au}},
   Chong Shen Ng\aff{1},
   Detlef Lohse\aff{1},
   Roberto Verzicco\aff{1,3,4},
   Dominik Krug\aff{1} \corresp{\email{d.j.krug@utwente.nl}}
  }

 \affiliation{\aff{1}Physics of Fluids Group, Max Planck UT Center for Complex Fluid Dynamics, Faculty of Science and Technology, and J.M. Burgers Centre for Fluid Dynamics, University of Twente, P.O. Box 217, 7500 AE Enschede, The Netherlands
 \aff{2} Department of Mechanical Engineering, University of Melbourne, Victoria 3010, Australia
\aff{3} Dipartimento di Ingegneria Industriale, University of Rome ‘Tor Vergata’,\\ Via del Politecnico 1, Roma 00133, Italy
\aff{4} Gran Sasso Science Institute - Viale F. Crispi, 7 67100 L'Aquila, Italy}

\newcommand{\ab}{\boldsymbol{a}}
\newcommand{\abc}{\boldsymbol{a}_C}
\newcommand{\eb}{\boldsymbol{e}}
\newcommand{\ub}{\boldsymbol{u}}

\newcommand{\vb}{\boldsymbol{v}}
\newcommand{\vbc}{\boldsymbol{v}_C}
\newcommand{\vbg}{\boldsymbol{v}_G}

\newcommand{\fb}{\boldsymbol{f}}
\newcommand{\rb}{\boldsymbol{r}}
\newcommand{\gb}{\boldsymbol{g}}
\newcommand{\pb}{\boldsymbol{p}}
\newcommand{\qb}{\boldsymbol{q}}
\newcommand{\xb}{\boldsymbol{x}}

\newcommand{\ut}{\hat{u}_{\theta}}
\newcommand{\rh}{\hat{r}}

\newcommand{\Str}{\mbox{\textit{Str}}}

\newcommand{\Fb}{\boldsymbol{F}}

\newcommand{\Lcalb}{\boldsymbol{\mathcal{L}}}

\newcommand{\mt}{\mathcal{T}}
\newcommand{\mts}{\tilde{\mathcal{T}}}

\newcommand{\phib}{\boldsymbol{\varphi}}

\newcommand{\zetab}{\boldsymbol{\zeta}}
\newcommand{\omb}{\boldsymbol{\omega}}

\newcommand{\Gal}{\textit{Ga}} 

\newcommand{\rms}{{\textrm{rms}}}

\newcommand\cf{cf.\ }
\newcommand\ie{i.e.\ }

\begin{document}

\maketitle

\begin{abstract} 
Rotational effects are commonly neglected when considering the dynamics of freely rising or settling isotropic particles. Here, we demonstrate that particle rotations play an important role for rising as well as for settling cylinders in situations when mass eccentricity, and thereby a new pendulum timescale, is introduced to the system. We employ two-dimensional simulations to study the motion of a single cylinder in a quiescent unbounded incompressible Newtonian fluid. This allows us to vary the Galileo number, density ratio, relative moment of inertia, and Centre-Of-Mass offset (COM) systematically and beyond what is feasible experimentally. For certain buoyant density ratios, the particle dynamics exhibit a resonance mode, during which the coupling via the Magnus lift force causes a positive feedback between translational and rotational motions. This mode results in vastly different trajectories with significantly larger rotational and translational amplitudes and an increase of the drag coefficient easily exceeding a factor two. We propose a simple model that captures how the occurrence of the COM offset induced resonance regime varies, depending on the other input parameters, specifically the density ratio, the Galileo number, and the relative moment of inertia. Remarkably, depending on the input parameters, resonance can be observed for centre-of-mass offsets as small as a few percent of the particle diameter, showing that the particle dynamics can be highly sensitive to this parameter. 
\end{abstract}


\section{Introduction}\label{sec:intro}
One of the striking characteristics of freely rising and settling particles at sufficiently large Reynolds numbers is the existence of non-vertical paths. This type of motion is a common characteristic for the dynamics of blunt bodies in general \citep{Ern2012} and is related to the presence of a (periodically) oscillating wake with vortex shedding, akin to that forming behind a fixed geometry \citep{gerrard1966,perry1982,williamson1996}. The phase regimes of freely rising or settling bodies are more complicated than their fixed counterparts, however, due to the inherent coupling between the motion of the body-fluid interface and the surrounding flow morphology. This results in a complex, often only quasi-periodic motion which generally is difficult to predict a priori. Therefore, a complete solution or model of these systems remains elusive in many cases such that new insights rely on empirical results and parameter studies based on experiments or numerical simulations. Understanding and modelling of particle dynamics is important in many fields, for instance to predict the the spread of (plastic) pollutants in the ocean \citep{sutherland2023}.

Properties of the paths and dynamics observed for buoyancy/gravity driven motion are determined by the strength of the coupling between particle and fluid. When this coupling is weak \citep{horowitz2006} or when the degrees of freedom of motion are limited \citep{williamson2004}, then the fluid response will be similar to that of a fixed geometry. On the contrary, regimes exist where particle kinematics are strongly affected by the fluid motion, and vice versa leading to alterations in flow morphology and particle trajectory and dynamics, \eg the shedding frequency, path amplitude, and most practically relevant; the drag. These changes to particle dynamics and kinematics are important for our understanding of e.g. the mixing behaviour in chemical reactors and in waste/resource extraction by flotation and sedimentation \citep{almeras2015mixing,chan2023bubble} or in natural processes such as sedimentation in rivers or oceans \citep{meiburg2010}, particle-turbulent interaction of falling snowflakes in the atmosphere \citep{nemes2017}, or as previously mentioned the spread of micro plastics in our oceans \citep{sutherland2023}.

Previous studies have often focused solely on the translational dynamics in relation to the particle-to-fluid density ratio, often disregarding body rotation, as noted by \citet{mathai2017}. In the present work, we primarily focus on the effects of rotational coupling. To this end, we vary the internal mass distribution of freely rising and settling two-dimensional (2D) cylinders by introducing a Centre-Of-Mass (COM) offset. This approach is motivated by recent work of \citet{will2021:COM}, where the COM of freely rising and settling spheres was varied experimentally. This study found that the rotational dynamics of the spheres are strongly affected by the internal mass distribution, which in turn strongly affects all other aspects of particle motion. Stronger rotational dynamics are also observed for anisotropic particles. However, in this case inertial (normal) forces on the particle also contribute a net torque, such that the driving in this case is not exclusively via the rotational-translational coupling introduced by the COM.
The observed phenomena, for the case of spheres, could be explained via the analogy to a simple driven harmonic oscillator and expressing the results in terms of a timescale ratio between the `pendulum' frequency, induced by the offset, and the driving frequency, which is determined by the vortex-shedding. This model captured several key features of the particle dynamics when COM offset was present. It further predicted additional dependencies on the particle-to-fluid density ratio $\Gamma$, the dimensionless Moment Of Inertia (MOI) of the particle $I^*$, and implicitly on the Galileo number \Gal, which is similar to the Reynolds number $Re$  as it is a  measure of the ratio between inertial to viscous forces but uses an a priori defined buoyancy velocity scale instead of a dynamical one, and governs the onset and transitions in between the various wake-structure topologies. However, due to experimental and physical constraints, the parameter space available in \citet{will2021:COM} is insufficient to test these conclusively. Therefore, we aim to systematically investigate these dependencies by means of numerical simulations of 2D cylinders with COM offset, to show that the underlying physics are similar between the 2D and the three-dimensional (3D) case, and that the results presented here can shed light on the remaining open questions.

The problem of freely rising or settling cylinders is an extension of the classical case of vortex-induced vibrations \citep{bearman1984,parkinson1989,govardhan2000,williamson2004}, where a cylinder is placed in a free stream with only limited degrees of freedom. The applied restrictions indicated that the degrees of freedom and, therefore, the amount of particle motion (tuned by body inertia, spring stiffness, and damping of the supports) strongly influence the wake structure and coupled dynamics. \citet{williamson1988} presented qualitatively similar results for a cylinder that was forced periodically in a free stream and classified the resulting wake patterns as a function of the driving amplitude and frequency. Building on this, the work by \citet{jeon2004} showed that the type of vortex-shedding depends on transverse and streamwise oscillations as well as on their relative phase. Similarly, for elastically mounted cylinders at subcritical Reynolds numbers ($\Rey \leq30$) the effects of forced rotations were examined in recent work by \citet{bourguet2023}, uncovering significant alterations in flow structure and amplitude of oscillation depending on $\Rey$ and rotational magnitude and frequency. For the case of freely rising and settling 2D cylinders, a critical mass density ratio ($\Gamma$), the ratio of particle to fluid mass density, was encountered, marking the threshold between reduced coupling at high $\Gamma$ where the particle dynamics and its wake barely influence each other. On the contrary, below this threshold, particles show large path amplitudes and substantial alterations in the wake vortex shedding frequency \citep{horowitz2006,horowitz2010vortex}. Similar density ratio related transitions in the regime of motion have also been observed for spheres \citep{Horowitz:2010,Auguste2018,will2021:MOI}. Following the same train of thought, the rotational moment of inertia was also investigated as a relevant parameter, governing the dynamics of rising and settling 2D cylinders in the numerical work by \citep{mathai2017}, as well as experimentally for spheres \citep{Mathai2018,will2021:MOI}. Due to these previous observations, we investigate the effects of $\Gamma$ and MOI separately and in combination with effects induced by a COM offset.

Before proceeding with the problem definition, caution is warranted when interpreting the results as the 2D assumption in this work effectively corresponds to the limiting case of particle motion for very long cylinders settling or rising in a three-dimensional (3D) environment. Beyond a certain Reynolds number, the flow will become inherently 3D, even for a cylinder of infinite length; for a fixed cylinder, this is found to occur around $\Rey \approx 190$ \citep{henderson1997,williamson1998series,aleksyuk2023}. Moreover, the cylinder length and associated end-effects play an important role \citep{inoue2008}, such that the motion of these cylinders becomes inherently 3D, showing both horizontal (azimuthal) cylinder rotation (around the vertical axis) and fluttering motion (around a horizontal axis) \citep{toupoint2019}. Therefore, we would like to preface this work by stating that the principle goal is not to perfectly predict dynamics of 3D cylinders settling or rising in a quiescent fluid but rather to better understand the physics underlying the behaviour resulting from different combinations of the four control parameters for both cylindrical and, more relevantly spherical geometries.

\subsection{Problem definition and equations of motion}\label{sec:problem_definition}
\begin{figure}
	\centerline{\includegraphics[width=1\textwidth]{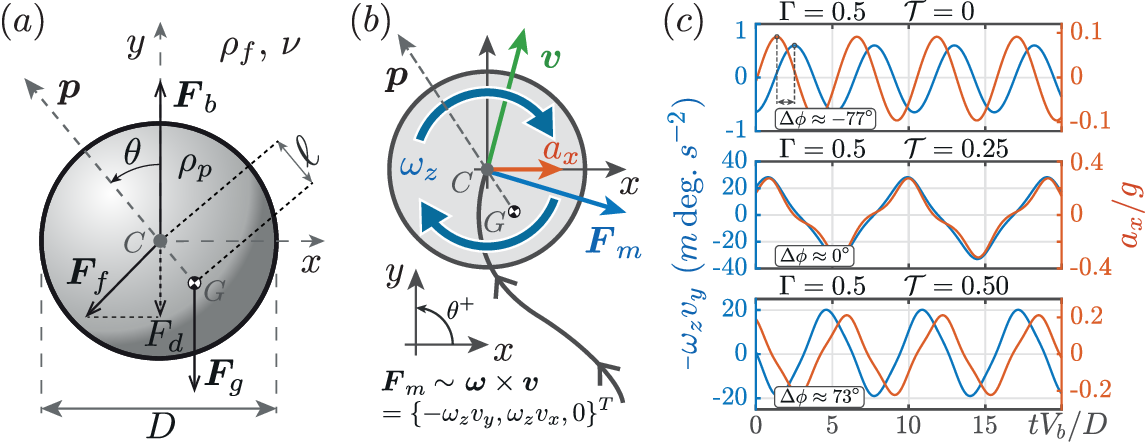}}
	\caption{({\it a\/}) Schematic of the cylinder with the centre-of-mass ($G$) displaced by a distance $\ell$ from the volumetric centre ($C$). The pointing vector $\pb$ is a unit vector in the direction from $G$ to $C$ and $\theta$ is the angle between $\pb$ and the vertical ($y$-direction). The forces acting on the body are buoyancy ($\Fb_b$) and the remaining fluid forces $\Fb_f$ (in $C$) and gravity $\Fb_g$ (in $G$). ({\it b\/}) Schematic depicting the direction of the Magnus lift force the horizontal component of which is used together with the horizontal particle acceleration to calculate the phase lag $\Delta \phi$. ({\it c\/}) Time signals of the horizontal component of the Magnus force ($\Fb_m\sim -\omega_z v_y$, blue) and horizontal particle acceleration ($a_x/g$, red) for three different offset cases, showing different phase lags. Note that, since $\langle v_y \rangle = \mathcal{O}(1)$, the value on the left $y$-axis is indicative of body rotation.}
	\label{fig:probDef}
\end{figure}
In this work, we concern ourselves with the dynamics and kinematics of freely rising and settling 2D cylinders in an otherwise quiescent, infinite fluid. The fluid phase has a constant mass density $\rho_f$ and a kinematic viscosity $\nu$. The motion of the cylinder is confined to move within the $xy$-plane. Here, the $y$-axis is anti-parallel to the gravity vector (which has magnitude $g$). Subscripts $x$ and $y$ are assigned to vector components in this plane. We denote particle position, velocity, and acceleration by $\boldsymbol{x}_p$, $\vb$, and $\ab$, respectively.

The particle (see schematic in figure \ref{fig:probDef}\,({\it a\/})) has a circular diameter $D$, and an effective mass $m_p$, mass density $\rho_p$, and a volume $\mathcal{V}$,  per unit length. The COM of the cylinder (designated with point $G$) is displaced by a distance $\ell$ from the geometric centre $C$. Subscripts $C$ and $G$ throughout will refer to these points. A unit pointing vector $\pb$ is defined between these points from $G$ to $C$. From here, we define the angles $\theta$ between $\pb$ and $\eb_y$ (the vertical unit vector), and $\theta_v$ between $\pb$ and $\vbc$ (the instantaneous particle velocity at the centre). The buoyancy force acts upwards through point $C$, while the gravitational force acts downwards through point $G$. The relevant velocity scale characterising this system is the buoyancy velocity, \ie $V_b=\sqrt{|1-\Gamma|gD}$.

The dynamics of freely rising and settling buoyancy-driven particles are governed by the linear and angular momentum balances. In the present work, all particle dynamics are for unconstrained motion, implying that the only two forces acting on the geometry are a body force due to gravity ($\Fb_g = -\rho_p\mathcal{V}g\eb_y$) and the force exerted by the fluid on the particle $\Fb_F$. For convenience we split this total fluid force into a contribution due to buoyancy and a time-varying part $\Fb_F = \Fb_b + \Fb_f$, where $\Fb_b = \rho_f\mathcal{V}g\eb_y$. Therefore, the conservation of linear momentum for the cylinder is given by
\begin{equation}
    \Gamma \dfrac{\textrm{d}\vbg}{\textrm{d}t} =  \dfrac{\Fb_f}{m_f} + (1-\Gamma)g\eb_y,
    \label{eq:KKT_translation}
\end{equation}
with $\Gamma =\rho_p/\rho_f$ the mass density ratio, and $m_f$ the mass of the fluid displaced by the particle. It is important to note that (\ref{eq:KKT_translation}) is independent of the COM offset. The angular momentum balance for a 2D cylinder with COM offset reads
\begin{equation}
I_C \dfrac{\textrm{d}^2 \theta}{\textrm{d} t^2} = T_f - \dfrac{1}{2}  m_p  \gamma D  (g\eb_y +\abc) \times \pb,
\label{eq:KKT_rotation}
\end{equation}
where the balance is constructed with respect to the \emph{geometric centre} of the particle. Here, $T_f$ is the fluid torque due to viscous stresses on the particle and the additional terms $(g\eb_y+\abc) \times \pb$ on the right-hand side results from the COM offset. In the current work we treat $I_C$, the moment of inertia of the particle around point $C$, as the governing parameter characterising the body's rotational inertia. With $I_G$ following from the parallel axis theorem: $I_G = I_C - m_p \ell^2$. Further, $\gamma$ denotes the ratio $\gamma = 2\ell/D$. The rotational dynamics described in (\ref{eq:KKT_rotation}) resemble those of a forced pendulum \citep{will2021:COM}, for which, when linearised, the natural frequency is given by
\begin{equation}
    f_p = \tau_p^{-1}\ = \dfrac{1}{\pi}\sqrt{\dfrac{\gamma g}{D I^*}},
    \label{eq:fp}
\end{equation}
with $I^* = I_C/I_\Gamma$, and $I_\Gamma = m_p D^2/8$ a reference value of a homogeneous cylinder with identical mass $m_p$. Therefore, the physically relevant range is $0 \leq I^* \leq 2$, i.e. $0 \leq I_C \leq 1/4 m_p D^2$ (all mass in the centre all mass on the edge, respectively). In our work, however, we consider cases beyond $I^* = 2$ to unravel the role of this parameter. We can further write (\ref{eq:KKT_rotation}) in dimensionless form by introducing a dimensionless timescale $\tilde{t} = t/\tau_v$, where $\tau_v = D/V_b$ is the vortex shedding timescale and therefore represents the typical timescale of the forcing in (\ref{eq:KKT_rotation}). Since the geometry is cylindrical and only shear forces contribute to the torque around $C$, we use viscous scaling to non-dimensionalise the viscous torque term as $T_f = \mu D L V_b\, T^*_f$ \citep{jordan1972,bouchet2006}. Here, $L$ is the length of the cylinder which is set to unity in the present work. On the other hand, since the contribution of the body acceleration is related to lateral, pressure induced, forcing we use inertial scaling in order to non-dimensionalise this term writing the particle acceleration as $\boldsymbol{a}_C \sim \boldsymbol{F}_f/m_p$, and using inertial scaling $||\Fb_f|| \sim \rho_f L D V_b^2$ to obtain $\abc  = V_b^2/(D \Gamma) \, \abc^*$. Applying this to \ref{eq:KKT_rotation} we obtain:

\begin{equation}
    \dfrac{\textrm{d}^2 \theta}{\textrm{d} \tilde{t}^2} = \cfrac{32}{\pi}\dfrac{1}{Ga \Gamma I^*}\, T^*_f  -  \dfrac{4\gamma}{I^*}\left(  \dfrac{1}{|1-\Gamma|} \eb_y +  \dfrac{1}{\Gamma} \abc^* \right) \times \pb.
    \label{eq:KK_Tstar}
\end{equation}
Here, the Galileo number is defined as $\Gal = \sqrt{|1-\Gamma|gD^3}/\nu$. The dimensionless prefactor $\gamma/(|1-\Gamma|I^*)$ in front of the pendulum term is proportional to the square of the ratio $\mt$ of the vortex shedding to the pendulum timescale as defined in \citet{will2021:COM}. For a 2D cylinder, this timescale ratio is equal to
\begin{equation}
    \mt = \dfrac{\tau_v}{\tau_p} = \dfrac{1}{\pi}\sqrt{\dfrac{\gamma}{|1-\Gamma| I^*}}.
    \label{eq:T}
\end{equation}
Note that the control parameter $\mt$ is solely dependent on the prescribed particle and fluid properties. It was shown by \citet{will2021:COM} that this control parameter governs the rotation dynamics of spheres with a COM offset. We will validate and expand upon this finding in the present work.\\

In \S\,\ref{sec:numerics} we will first outline the numerical approach used to obtain the results as well as the data processing applied to the data set. Then, in \S\,\ref{sec:Gamma} -- \S\,\ref{sec:Galileo}, our results are presented and discussed. This is split into four sections, where  \S\,\ref{sec:generalCOM} contains a general discussion on effects of COM offset, the resonance mechanism, and the effect of fluid inertia for freely rising cylinders. Next, in \S\,\ref{sec:Gamma} the role of the particle-to-fluid density ratio and how it affects COM-induced phenomena is discussed along with the differences between rising and settling, followed by a discussion on MOI effects in \S\,\ref{sec:moi} and Galileo number effects in \S\,\ref{sec:Galileo}. Finally, in \S\,\ref{sec:conclusion} the primary results and findings are summarised.

\section{Numerical framework}\label{sec:numerics}
\subsection{Fluid phase}
The incompressible Navier--Stokes equations describe the flow of an unbounded Newtonian fluid around the particle and satisfy the boundary conditions at the body surface and infinity.  An approximate computational strategy to model this configuration is achieved by solving the Navier--Stokes equations on a finite size domain in a moving reference frame attached to the body. For a perfectly circular particle, this reference frame does not need to rotate due to the body's inherent symmetry. A co-moving frame also allows for a configuration where the gravity vector is directed towards the outlet such that the wake can gently leave the domain without disturbing the particle dynamics. The incompressible Navier--Stokes equations are non-dimensionalised with $V_b$ and $D$. For the translating frame, the momentum and continuity equations are given by \citep[see e.g.][]{mougin2002,jenny2004}
\begin{subequations}
    \begin{equation}\label{eq:NS_Momentum}
    \frac{\partial\ub}{\partial t} + \nabla \cdot [\ub(\ub-\vbc)] = -\nabla p + \dfrac{1}{\Gal} \nabla^2 \ub +\fb, 
    \end{equation}
\begin{equation}\label{eq:NS_Continuity}
    \nabla\cdot \ub=0, 
\end{equation}  
\end{subequations}
with $\ub$ the fluid velocity vector, $p$ the kinematic pressure and $\fb$ the boundary forcing from the immersed boundary method (IBM) that enforces the no-slip condition (described in detail \S\,\ref{sec:num_method}). Note that the hydrostatic component is absent from $p$ as it is explicitly added to the forces acting on the cylinder.

The velocity at the inflow is set to zero to simulate an asymptotically quiescent fluid. At the outflow, a convective boundary is imposed \citep[see e.g. ][]{kim2006}. The side walls of the two-dimensional domain are periodic. The domain size is set to $60D$ in the gravity direction and $20D$ in the transverse direction, which was found to be sufficiently large to avoid box size effects. The grid spacing was kept constant within a square of size $2D$ adjacent to the cylinder. Outside this domain, the mesh spacing expands linearly towards the edges of the domain. A heat equation is solved to smoothen the mesh transition from the fine constant spacing surrounding the cylinder to the linearly expanding mesh, which is required to avoid numerical artefacts, and the final grids employed for the various cases are presented in table \ref{tab:grids}. These grids were chosen such that the particle boundary layer was resolved by three to four grid points. Tests confirmed that halving the respective grid spacing did not alter the overall statistics. 

\begin{table}
\centering
\begin{tabular}{ccc}
$\Gal$ & $D/\Delta x$  & $N_x\times N_y$ \vspace{2pt} \\
50-200   &  50  & $420\times960$\\ 
500   &  84 & $520\times1104$ \\
700   &  95 & $560\times1152$\\
2000 & 161 & $700\times1440$
\end{tabular}
\caption{Overview of the grids. The first column denotes the Galileo number $\Gal$. The second column represents the number of grid points per diameter of the cylinder. The third column is the grid resolution for the fluid phase.}
\label{tab:grids}
\end{table}

\subsection{Numerical method}\label{sec:num_method}
The numerical scheme closely follows the immersed boundary projection method originally developed by \citet{taira2007} and \citet{lacis2016} of which a brief overview is presented. We solve \eqref{eq:NS_Momentum} and \eqref{eq:NS_Continuity} on a staggered grid, where the spatial gradients are computed using a conservative second-order central finite difference scheme.
The non-linear term of \eqref{eq:NS_Momentum} is advanced in time via the explicit second-order Adams--Bashforth scheme and the viscous terms via the second-order implicit Crank--Nicolson scheme:

\begin{subequations}
  \begin{align}\label{eq:NSdiscretised}
		\begin{split}
		\frac{\ub^{n+1}-\ub^n}{\Delta t} &+ \frac{3}{2}\hat{N}(\ub^n,\vbc^n) - \frac{1}{2} \hat{N}(\ub^{n-1},\vbc^{n-1}) = - \hat{G}\phib^{n+1/2} \\ & -\hat{G}\pb^n+ \frac{1}{2\Gal}\hat{L} (\ub^{n+1} +\ub^n) + \hat{H} \fb^{n+1/2},  
	\end{split}  \end{align}
\begin{equation} \label{eq:constraints}
\textrm{where}\quad \hat{D}\ub^{n+1}=0 \quad \text{and} \quad \hat{E} \ub^{n+1}=\vb_C^{n+1}+\omb^{n+1}\times \Lcalb.
\end{equation}
\end{subequations}
Here, $\hat{N}(\ub,\vb_C)$ denotes the non-linear operator, $\hat{G}$ the gradient operator, $\hat{L}$ the Laplace operator, $\hat{D}$ the divergence operator, $\hat{H}$ the spreading (regularisation) operator, $\hat{E}$ the interpolation operator, $\phib^{n+1/2}$ the discrete incremental pressure, $\pb^n$ the discrete pressure, $\Lcalb$ the Lagrangian marker coordinates with respect to geometric centre and $\fb^{n+1/2}$ the discrete analogue of $\fb$ in \eqref{eq:NS_Momentum}. The interpolation and regularisation matrices $\hat{E}$ and $\hat{H}$, respectively make use of a discrete three-point $\delta$-function \citep{roma1999}. 

For the particle, the non-dimensional Newton--Euler equations are advanced in time via
\begin{equation}\label{eq:discreteNewtEul}
\dfrac{1}{\Delta t} (\vbc^{n+1}-\vbc^n) =C_B(N_B\tilde{\fb}^{n+1/2} + \Delta\qb_B) + \gb^n - \zetab^n,
\end{equation}
with
\begin{equation}
C_B = \dfrac{4}{\pi\Gamma} \begin{bmatrix}
1 & 0 & 0 \\ 0 & 1 & 0 \\0 & 0 & 8/I^*\end{bmatrix},\quad
 N_B  =
- \begin{bmatrix}
1 & \ldots & 1 & 0 & \ldots & 0\\
0 & \ldots & 0 & 1 & \ldots & 1\\
-\mathcal{L}_{y_1} & \ldots & -\mathcal{L}_{y_{n}} & \mathcal{L}_{x_1} & \ldots & \mathcal{L}_{x_{n}} 
\end{bmatrix},
\end{equation}
 $\Delta\qb_B\equiv Q (\ub^n -\ub^{n-1})/\Delta t$ and $Q$ the matrix that interpolates the velocity inside the cylinder \cite[\cf][]{kempe2012}. The time step is limited with $\textrm{CFL}=0.4$.
 
 Vector $\gb^n$ contains the buoyancy force and the torque induced by the particle weight.
 The Newton equation in \eqref{eq:KKT_translation} is solved with respect to the geometric centre and we make use of $\vbc= \vbg + \gamma/2\,\omb\times\pb$ for the transformation of \eqref{eq:KKT_translation}. Additionally, the equation of angular conservation is solved with respect to the geometric centre (see \S\,\ref{sec:intro}) yielding an additional $\abc\times \pb$ term. The terms from the latter contributions are collected in $\zetab^n$. Finally, we have for $\gb^n$ and $\zetab^n$:

\begin{equation}
\gb^n \equiv \begin{Bmatrix} 0\\  \textrm{sgn}(1-\Gamma)/\Gamma\\ - 4\gamma/I^*  \sin\theta^n \end{Bmatrix},\quad 
\zetab^n \equiv \dfrac{\gamma }{2\Delta t} 
\begin{Bmatrix}
\omega^n\cos\theta^n - \omega^{n-1}\cos\theta^{n-1}   \\ \omega^n\sin\theta^n - \omega^{n-1}\sin\theta^{n-1} \\ 
(a_x^n \cos\theta^n + a_y^n\sin\theta^n) / I^* 
\end{Bmatrix},
\end{equation}
with $a_x^n\equiv v_x^n-v_x^{n-1}$ and $a_y^n\equiv v_y^n-v_y^{n-1}$, discretised components relating to $\abc$, respectively.
 
Equation \eqref{eq:NSdiscretised} together with constraints  \eqref{eq:constraints}, and \eqref{eq:discreteNewtEul}  can be rewritten as
\begin{equation}\label{eq:NS_discr_matrix_form}
\begin{bmatrix}
A & 0 & G & E^T \\ 0 & I_B & 0 & N_B \\ G^T & 0 & 0  & 0 \\ E & N_B^T & 0 & 0
\end{bmatrix} \begin{Bmatrix} \qb^{n+1} \\ \vbc^{n+1} \\ \phib^{n+1/2} \\  \tilde{\fb}^{n+1/2} \\  \end{Bmatrix} = \begin{Bmatrix} \rb^n \\ \rb_B^n \\\boldsymbol{0}\\ \boldsymbol{0} \end{Bmatrix},
\end{equation}
with $\qb^{n+1}=R\ub^{n+1}$, $E = \hat{E}R^{-1}$, diagonal matrix $R\equiv[\Delta y_j,\, 0; 0,\, \Delta x_i]$, $A=\hat{M}\left[I/\Delta t -\hat{L}/(2\Gal)\right]$, and $\rb^n$, $\rb_B^n$ containing the explicit terms. 
We approximate the inverse of $A$ as $A^{-1}\approx B= \Delta t M^{-1}$   \cite[\cf \S\,3]{lacis2016}, with $M=\hat{M}R^{-1}$, $\hat{M}\equiv[(\Delta x_i+\Delta x_{i-1})/2,\,  0; 0,\, (\Delta y_j+\Delta y_{j-1})/2)]$ a diagonal matrix. The solution procedure of \eqref{eq:NS_discr_matrix_form} is performed via a block-LU decomposition following a three step procedure
\begin{subequations}
\begin{equation} \label{eq:aq_b_is_r}
      \begin{bmatrix} A & 0\\ 0 & I_B \end{bmatrix}
      \begin{Bmatrix} \qb^* \\ \vbc^* \end{Bmatrix} = 
      \begin{Bmatrix} \rb^n \\ \rb_B^n  \end{Bmatrix},
\end{equation}
  \begin{equation}\label{eq:q*sol}
      \begin{bmatrix}
      G^TBG & G^TB E^T \\ EBG & EBE^T + N_B^T I_B^{-1}N_B
      \end{bmatrix} 
      \begin{Bmatrix}
     \phib^{n+1/2} \\   \tilde{\fb}^{n+1/2}
      \end{Bmatrix}
      = \begin{Bmatrix}G^T\qb^* \\ E \qb^* + N_B^T\vbc^* \end{Bmatrix},
  \end{equation}
  \begin{equation}\label{eq:qbn+1}
            \begin{Bmatrix} \qb^{n+1} \\ \vbc^{n+1} \end{Bmatrix}
             = \begin{Bmatrix} \qb^* \\ \vbc^* \end{Bmatrix}
             -\begin{Bmatrix}B G\phib^{n+1/2}+ B E^T \tilde{\fb}^{n+1/2} \\ I_B^{-1} N_B\tilde{\fb}^{n+1/2}   \end{Bmatrix}.
  \end{equation}
\end{subequations}
Here, $\qb^*$ in \eqref{eq:aq_b_is_r} is solved for via a well tested factorisation procedure \cite[see e.g.][]{verzicco1996}. The solution of  $\phib^{n+1/2}$  and $\tilde{\fb}^{n+1/2}$ are obtained via the PETSc library \citep{petsc-efficient,petsc-user-ref} using the algebraic multigrid method BoomerAMG as the preconditioner and the general minimum residual method (GMRES) to solve \eqref{eq:q*sol}. This combination of solvers was found to be robust and converge within 12 to 17 iterations depending on the selected grid size and time step.
A relative tolerance was set to $10^{-13}$, to obtain solutions that satisfy the divergence free condition and no slip condition up to the limit of double-precision calculations.
Once the solution vector is found, we update the pressure field \citep[\cf][]{verzicco1996}
\begin{equation}
    \pb^{n+1} = \pb^{n} + \phib^{n+1/2} - \dfrac{\Delta t}{2\Gal}\hat{L}\phib^{n+1/2}.
\end{equation}

The additional solving routines for equations \eqref{eq:q*sol} and \eqref{eq:qbn+1} were tested to ensure that they yield machine precision solutions satisfying the divergence free, and no-slip condition (defined in \ref{eq:constraints}). The overall solution procedure was found to provide first-order convergence rate in the $L_2$ norm for the velocity field and first-order in time (owing to the approximation of $A^{-1}\approx \Delta t M^{-1}$). Multiple validations for fixed, and freely rising cylinders showed good agreement with previous numerical and experimental studies (see Appendix \ref{app:eomparticle}). The method was found to be stable, even for density ratios as low as $\Gamma=0.001$. This stability is inherent to the  coupling between the particle and the flow via the matrix definitions defined in \eqref{eq:qbn+1}. This means that we may use explicit finite differences for the predictor velocity $\vb_C^*$ in $\rb^n_B$, since the corrector velocity  $\vb_C^{n+1}$ is solved for simultaneously with $\qb^{n+1}$.

\subsection{Data set and processing} \label{app:dataset}
In this work, a total of 938 cases have been simulated for different combinations of the four control parameters: $\mt$, $\Gal$, $\Gamma$ and $I^*$. The main goal is to investigate the effect of the COM offset in combination with the other parameters. For this, we varied $\mt$ between 0 and a maximum of 0.6, $\Gal$ in the range between 50 up to 2000, $\Gamma$ between 0.001 and 5, and $I^*$ from 0.5 to 16. Compiled input and output parameters for all cases in our data set are included in the supplementary materials.

All results presented in this study are obtained after a statistically steady state has been reached. To ensure this, first, a moving average is computed of the time trace of the vertical velocity with an averaging window much larger than a single period of the typical fluctuations. This processed signal is compared to the terminal velocity of that case (determined by the average of the last 10\% of the time signal). The initial transient is considered to have ended once the filtered time signal deviates less than 5\% from this terminal velocity. For $\Gal=200$, this typically is the case after a transient time of $60 D/V_b$, which is short compared to the total average simulation time of $(1.9\times 10^3\,  D/V_b)$.

A number of different properties are derived from the simulations to characterise kinematics and dynamics of the particle path and the surrounding flow field. In the following, we describe the procedures used to extract these in detail.

The frequency $f$ of the horizontal path oscillations is determined by the peak of the power spectrum of $v_x(t)/V_b$, for which we applied local peak fitting in order to increase the accuracy of the estimated $f$. In the case of multiple peaks, the most prominent one is used in subsequent analysis and data visualisation. Some specific cases featuring multiple peaks are discussed in \S\,\ref{sec:Special_cases}. The obtained values of $f$ were cross-checked with an autocorrelation analysis of $v_x(t)/V_b$, which was found to yield almost identical results in all cases.

Due to the intrinsic unsteady and non-regular motion of these bodies, some additional processing is required to obtain oscillation amplitudes of the particle rotation and translation due to drift present in the time signals of $\boldsymbol{x}_p(t)$ and $\theta$. The reference $\theta = 0$ is either defined by the direction of the offset or by the initial orientation in the case of zero offset without loss of generality. To correct for the slow drift present for some of the cases, we employ a moving averaging filter on the signal with a window size of approximately $1/f(Ga,\Gamma,\mt,I^*)$, or one full oscillation time. Thus, we obtain a `centre-line' ($\boldsymbol{x}_{cl}(t)$, with horizontal mean drift velocity $v_d = \langle | \textrm{d} \boldsymbol{x}_{cl,\, x} /\textrm{d}t |\rangle$, documented in the supplementary data) which is subtracted from the actual position and orientation time signal to remove any low frequency effects. The absolute value of the signal processed this way is used to determine a list of the individual peak amplitudes ($A$) for the path and ($\theta$) for the rotational oscillations, the mean of which is denoted by $\hat{A}$ and $\hat{\theta}$, respectively. Note that, as a consequence, this can mean that the particle does not exhibit rotational oscillations around $\theta = 0$ (where $\pb$ is pointing upwards). Instead, especially for small offsets we observed a behaviour where $\theta$ might drift away from $\theta = 0$ followed by a large rotation back to the reference state when the rotational amplitude becomes large.

The phase lag $\Delta \phi$ between the horizontal component of the Magnus lift force $\Fb_m$ and the horizontal body acceleration $\ab_x$ is calculated via cross-correlation of these quantities. Similarly,  $\Delta \psi$ denotes the phase lag between the angular acceleration $\alpha$ and the fluid torque $T_f$. The lag obtained from the cross-correlation is divided by the length of an oscillation period $1/f$ and then expressed as a phase angle ranging from -180$^\circ$ to 180$^\circ$. The respective components that define $\Delta \phi$ are illustrated in figure \ref{fig:probDef}({\it b\/}). Figure  \ref{fig:probDef}({\it c\/}) provides three examples of signals with varying $\mt$ showing a negative, zero and positive value of $\Delta \phi$.

\section{General effect of the COM offset on dynamics and kinematics}
\label{sec:generalCOM}

\subsection{Particle kinematics and wake structures}
\label{sec:qualitative_kinematics_wake_structures}
\begin{figure}
	\centerline{\includegraphics[width=1.0\textwidth]{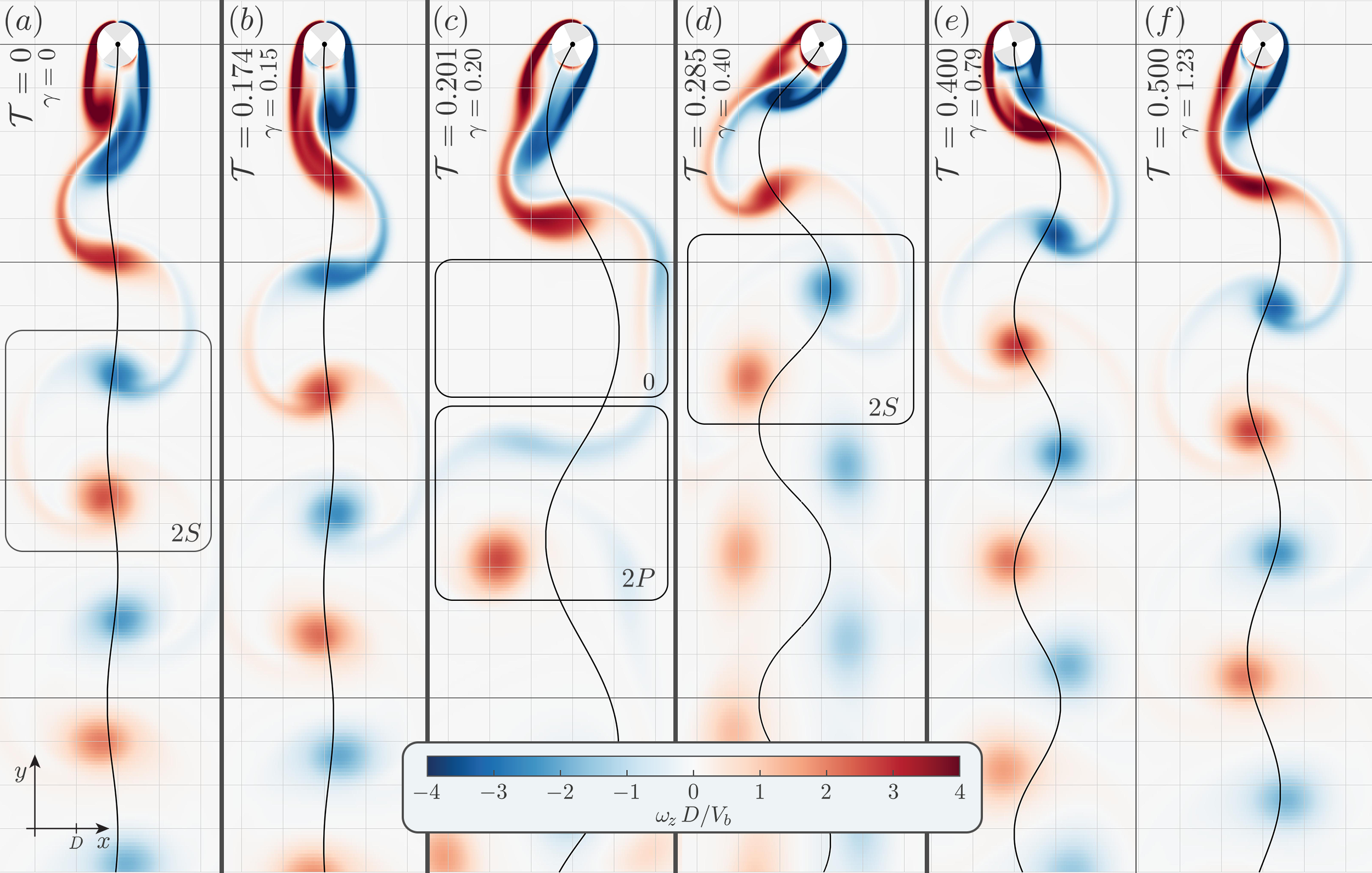}}
	\caption{Snapshots of particle trajectories and wake structures of rising cylinders with Galileo number $\Gal=200$ and density ratio $\Gamma=0.5$ for six different centre-of-mass offsets $\gamma$ (and $\mt$) ({\it a--f\/}). The offset increases from left to right as indicated by the listed parameters in the top left of each subfigure. Particle trajectories are indicated by the black lines, the grid spacing has dimensions of the particle diameter $D$. Coloured contours represent the normalised vorticity field ($\omega_zD/V_b$). }
	\label{fig:kinematics}
\end{figure}

We present figure \ref{fig:kinematics} to give an impression of how the wake patterns and particle kinematics change in the presence of COM offset. These snapshots display the non-dimensionalised fluid vorticity field ($\omega_z=\partial_y u_x-\partial_x u_y$) along with particle tracks (black lines) for six different COM offsets in the range $\gamma\in[0,\, 1.23]$ (increasing from left to right). All cases here are for $\Gal = 200$, $\Gamma = 0.5$, and $I^* = 1$.

The cylinder with zero COM offset in figure \ref{fig:kinematics}{\it a\/} is seen to rise almost straight, with regular vortex shedding occurring at double the frequency of the path oscillations. This vortex pattern, where two single vortices of opposite vorticity are shed during a single oscillation cycle, is the so-called ``2S'' mode \citep{williamson1988}. No visible effect of the COM offset is observed for cases up to $\mt = 0.174$ (figure \ref{fig:kinematics}{\it b\/}), but beyond this value, \eg $\mt = 0.201$ shown in panel figure \ref{fig:kinematics}{\it c\/}, remarkably different kinematics are encountered. Both amplitude and wavelength of the path oscillations are significantly larger in this case, and the wake now exhibits an irregular vortex shedding pattern as can be seen in \textcolor{black}{supplementary video 1}. For this case it is observed that the wake structure intermittently switches between several modes: (i) path oscillations with no significant shedding events (denoted by $0$ in figure \ref{fig:kinematics}), (ii) one single vortex pair as in the 2S-regime per oscillation, or (iii) four vortices per oscillation cycle; in two pairs of two shed when the body is changing direction; the so called ``2P''-mode. Note that these different modes appear to alternate without any noticeable long timescale pattern. This chaotic shedding pattern occurs for cases close to what we will call ``resonance'', where the rotational forcing induced by the path oscillations occurs at the same frequency as the inherent pendulum timescale. For even higher values of $\mt$ beyond resonance, represented by $\mt = 0.285$ in panel  figure \ref{fig:kinematics}{\it d\/}, we observe that the large amplitude path oscillations persist albeit with a reduced wavelength. Further, the vortex shedding returns again to an unperturbed 2S mode now with staggered vortex cores due to the strong path oscillations. Finally, with even larger offsets, figure \ref{fig:kinematics}({\it e,\,f\/}), the amplitude of the path oscillations begins to gradually reduce, returning to a state very much like that for the zero offset case (see \textcolor{black}{supplementary video 1} for $\mt >0.3$ cases). The results shown here are representative of the $\Gamma$-range where the resonance phenomenon is present. In the following, we will evaluate how this resonance behaviour depends on all of the other governing parameters.

\subsection{On the importance of fluid inertia}
\label{sec:fluidInertia}
\begin{figure}
	\centerline{\includegraphics[width=1\textwidth]{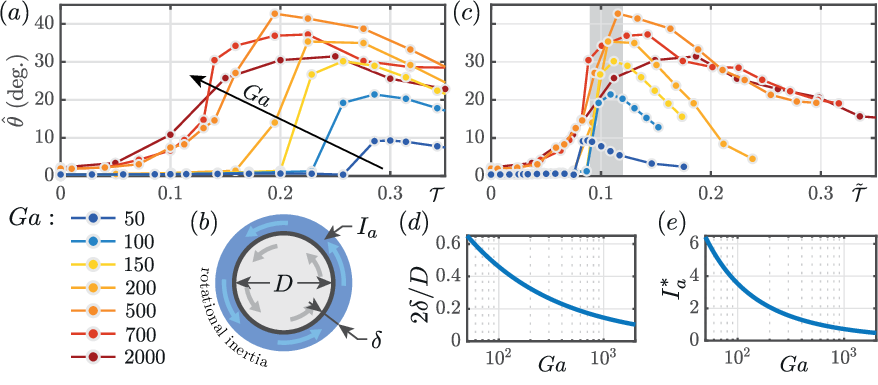}} 
	\caption{ ({\it a\/}) Mean rotational amplitude $\hat{\theta}$ as a function of Galileo number versus the timescale ratio $\mt$, here $\Gamma = 0.6$ and $I^* = 1$. ({\it b\/}) Schematic showing the parameters of the fluid inertia model. ({\it c\/}) $\hat{\theta}$ for the same cases as in ({\it a\/}) plotted against the modified timescale ratio $\mts$, which includes the effects of a Galileo number dependent added fluid inertia as per \eqref{eq:Tstar}. ({\it d\/}) Thickness of the fluid inertia layer $\delta$ and ({\it e\/}) added inertia as a function of $\Gal$, based on empirical collapse of the data.}
	\label{fig:fluidInertiaEffects}
\end{figure}

In order to investigate the effect of fluid inertia, we first consider the mean rotational amplitude ($ \hat{\theta} $) for a constant density ratio ($\Gamma = 0.6$) as a function of the timescale ratio $\mt$ as shown in figure \ref{fig:fluidInertiaEffects}({\it a\/}). Focusing initially on the case $\Gal = 200$ (corresponding to figure \ref{fig:kinematics}), we see that at $\mt = 0$ the rotational amplitude is small $\hat{\theta}  = 0.4^\circ$. Introducing a small amount of offset results in a marginal increase in this amplitude up to $\hat{\theta}  = 1.4^\circ$ at $\mt = 0.16$. However, around $\mt = 0.2$, there is a strong increase reaching a maximum amplitude of more than $\hat{\theta}  = 35^\circ$ at $\mt = 0.225$. This rapid increase is associated with the resonance phenomenon that was already visible in figure \ref{fig:kinematics}({\it c\/}). Beyond this point, the amplitude decreases gradually with increasing $\mt$. 

When comparing results across different $\Gal$ numbers, the same characteristic behaviour is observed for all cases in figure \ref{fig:fluidInertiaEffects}({\it a\/}). However, the value of $\mt$ at which resonance appears (marked by the steep increase in $\hat{\theta}$) is consistently shifted towards higher values as $\Gal$ decreases. Such a variation with $\Gal$ is not surprising since the definition of $\mt$ does not incorporate any viscous effects. However, for low values of $\Gal$ one would expect the Stokes layer surrounding the particle to contribute significantly to the total rotational inertia of the body and thereby also to modify the particle pendulum timescale. 
We can account for this effect by additionally including the rotational inertia $I_a$ resulting from the Stokes layer with thickness $\delta$ as illustrated in figure \ref{fig:fluidInertiaEffects}({\it b\/}) in our analysis. As a result, the modified pendulum frequency and timescale ratio of the system become
\begin{subequations}
\begin{equation}\label{eq:fpstar}
    \tilde{f}_p = \dfrac{1}{\pi}\sqrt{\dfrac{\gamma g}{D(I^* + I^*_a/\Gamma)}},
\end{equation}  
and 
  \begin{equation}\label{eq:Tstar}
    \mts  = \dfrac{1}{\pi}\sqrt{\dfrac{\gamma}{|1-\Gamma| \left( I^* + I^*_a/\Gamma \right) }},
\end{equation}
\end{subequations}
respectively. Here $I^*_a$ is the dimensionless fluid inertia defined as $I^*_a \equiv  8I_a/( m_f D^2)$, the ratio of the Stokes layer's rotational inertia to that of the displaced fluid. The total rotational inertia is thus given by $I^* + I^*_a\Gamma$. We assume that the thickness of this Stokes layer scales as $\delta \sim 1/\sqrt{\Gal}$ \citep{williamson1998series,schlichting2003,Mathai2018}, which for a cylinder leads to 
\begin{equation}
    I^*_a(\Gal) = \dfrac{8c_1}{\sqrt{\Gal}} + \dfrac{24c_1^2}{\Gal} + \dfrac{32 c_1^3}{\Gal^{3/2}} + \dfrac{16 c_1^4}{\Gal^2},
    \label{eq:I_star_a}
\end{equation}
with $c_1$ as the only free parameter. We find that choosing $c_1 = 2.3$ results in a reasonable collapse of the resonance regime for different $\Gal$ when plotting $\hat{\theta} $ against $\mts$ as shown in figure \ref{fig:fluidInertiaEffects}({\it c\/}). The corresponding thickness of the Stokes layer and magnitude of the added fluid inertia as a function of $\Gal$ are provided in figures \ref{fig:fluidInertiaEffects} ({\it d,\,e\/}), respectively. For $\Gal = 200$, the thickness of the fluid layer is approximately 0.33 particle radii and the rotational inertia amounts to about 2 times that of the displaced fluid. Beyond $\Gal = \mathcal{O}(10^3)$, the value of $I^*_a$ changes much more slowly, explaining the weak $Ga$ dependence at higher $\Gal$ observed in figure \ref{fig:fluidInertiaEffects}({\it a\/}) as well as in previous work on spheres \citep{will2021:COM}. Note, however, that $I_a^*$ is still 0.72 of the displaced fluid mass at $\Gal = 1000$ for cylinders and therefore by no means negligible. We performed an estimate of the history torque to confirm that the obtained values for $I_a^*$ are realistic. A complete discussion of this for both cylinders and spheres is provided in Appendix \ref{app:theor.anal.istar}.

\subsection{Who's driving?}
\label{sec:driving}
\begin{figure}
	\centerline{\includegraphics[width=1.0\textwidth]{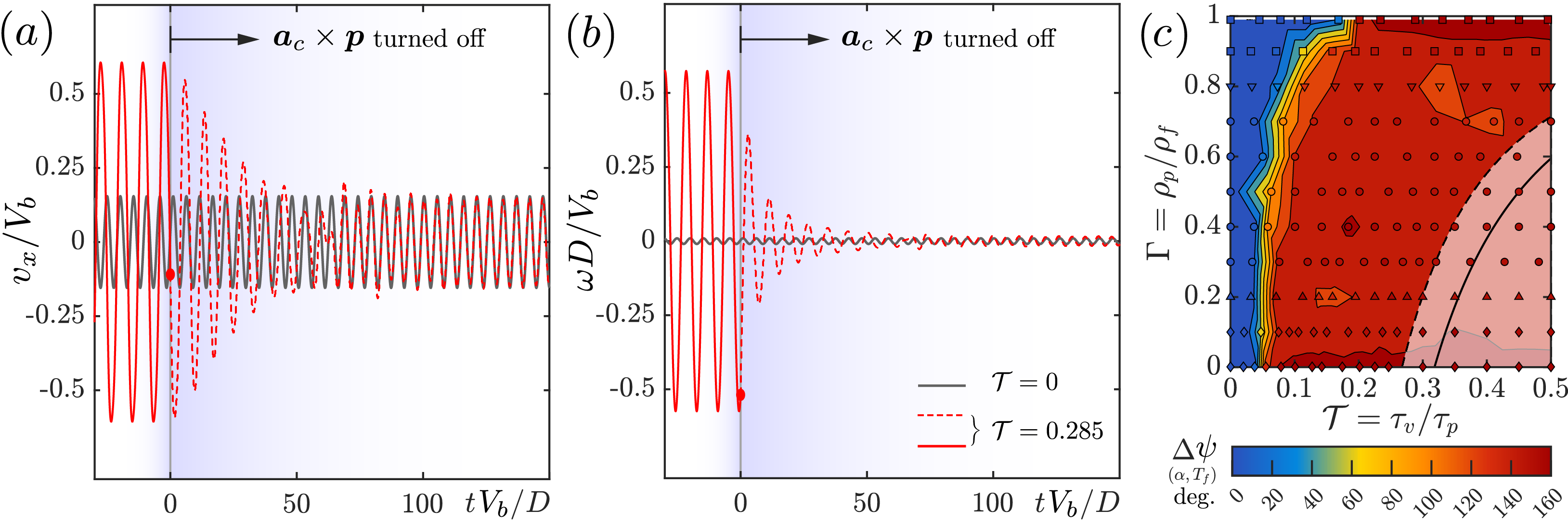}}
	\caption{ ({\it a,\,b\/}) Results for $Ga = 200$ and $\Gamma = 0.5$ for two cases; one without offset (grey line) and one with offset (red line). ({\it a\/}) Dimensionless horizontal velocity of the cylinder ($v_x$) and ({\it b\/}) dimensionless rotation rate ($\textrm{rad.}$) versus dimensionless time. During these runs at $t =0$ the $\abc \times \pb$ term for (\ref{eq:KKT_rotation}) is turned off, showing that in absence of this coupling term the dynamics of particles with offset almost completely revert back to those of particles without offset. ({\it c\/}) Phase lag $\Delta \psi$ between the rotational particle acceleration ($\alpha$) and the viscous torque ($T_f$).}
	\label{fig:acxp}
\end{figure}
When considering the right hand side of (\ref{eq:KKT_rotation}), there are two potential drivers of the rotational motion, the viscous torque $T_f$ and the translational-rotational coupling term $\abc \times \pb$, the latter being a consequence of the COM offset. Here, we will investigate their respective role with respect to the resonance behaviour. We know from the analysis in \S\ref{sec:fluidInertia} that the maximum in rotational amplitude is related to resonance between the vortex shedding timescale $\tau_v$ and the pendulum timescale $\tau_p$. However, both the viscous and translational driving will occur at the vortex shedding frequency, such that a distinction of their effects is not possible on this basis only. Answering the question of `who's driving' also provides insight in the effectiveness of COM offset in specific regimes of motion.

In order to untangle the effects of both contributions, simulations were performed where, after a statistically steady state had been reached, the ($\abc \times \pb$)-term was turned off in the integration of (\ref{eq:KKT_rotation}). In figure \ref{fig:acxp} ({\it a\/}), the horizontal component of the particle velocity $v_x$ is shown as a function of time for these runs at  $\Gal = 200$, $\Gamma = 0.5$, and $\mt = 0$ (grey line) and $\mt = 0.285$ (red line). At $t = 0$, the coupling term $\abc \times \pb$ is turned off for the case with offset. Figure \ref{fig:acxp} ({\it b\/}) displays the rotation rate $\omega$ for the same simulations. These results clearly indicate that as the coupling-term is turned off, the particle dynamics return to those of the case without offset. Note here that the pendulum term ($~\eb_y \times \pb$) is still present for $t\geq 0$, but evidently it has no effect without translational driving of the rotational dynamics. Therefore, we conclude that the rotational resonance phenomenon is linked to the translational coupling. As a consequence, we expect COM offset to have no impact on particle dynamics for cases where no horizontal path oscillations (i.e. no horizontal accelerations) are present, e.g. at low $Ga$. This also suggests that the resonance behaviour might also be triggered by outside periodic forcing, as would be present in a turbulent flow environment. It would be interesting to study how the settling/rising velocities of low Galileo number bodies with COM offset are affected in turbulence via this mechanism.

On the role of $T_f$, it is further instructive to consider the phase lag $\Delta \psi$ between $T_f$ and the rotational acceleration $\alpha$, which is shown in figure \ref{fig:acxp} ({\it c\/}) for the full range for $\Gamma$ and $\mt $ at $\Gal = 200$. For zero or very small offsets, $T_f$ is driving the (weak) rotational dynamics as evidenced by $\alpha$ and $T_f$ being close to in phase. However, as the offset increases towards resonance and beyond, $\Delta \psi$ switches swiftly to values close to 180$^\circ$, such that the viscous torque predominantly acts as damping in these cases. In essence, these trends also hold for higher $\Gal$. However, the dynamics become somewhat more chaotic at higher $\Gal$, as will be shown in \S \ref{sec:Galileo}, resulting in slightly lower values of $\Delta \psi$ on the order of $120$ to $150^\circ$.

\section{The effects of density ratio on COM offset}\label{sec:Gamma}
\subsection{Frequency of oscillation}\label{sec:freq_of_oscillation}
\begin{figure}
	\centerline{\includegraphics[width=1.0\textwidth]{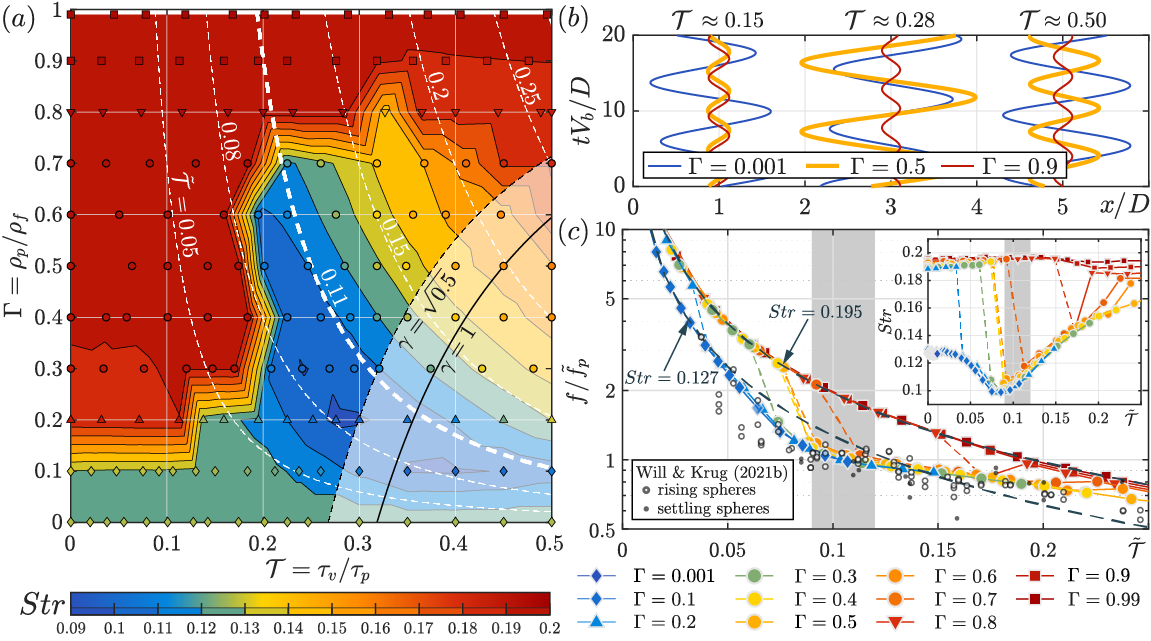}}
	\caption{Results for the path oscillation frequency ($f$) of rising particles at $Ga=200$ as a function of $\mt$ and $\Gamma$. 
	In ({\it a\/}), the marker colour indicates the dimensionless frequency ($Str = fD/V_b$) according to the colour bar provided below. The marker type indicates the different regimes in terms of the resonance behaviour discussed in the following. 
        The isocontours are based on a linear interpolation of the data. Dashed white lines represent isocontours of $\mts$, the timescale ratio including effects of fluid inertia.
	({\it b\/}) Horizontal particle position over the cylinder diameter ($x/D$) as a function of dimensionless time grouped in three values of $\mt$ for three values of the $\Gamma$ as indicated by the line colours showing characteristic behaviour for each.
	({\it c\/})  Ratio of the frequency ($f$) of the path oscillations over the pendulum frequency $f_p$ vs. the timescale ratio $\mts$. Here the marker colour indicates $\Gamma$ as listed in the legend below the figure. The two dashed black lines show a constant value of $Str$. Both of these show the collapse of COM and $\Gamma$ effects in terms of this parameter. The grey shaded region in this figure indicates the frequency lock-in regime. We further see that the results also collapse with the results from spheres with COM offset \citep{will2021:COM} shown as black symbols. The inset of the figure shows the same data as ({\it a\/}) plotted as $Str$ vs. $\mts$.
	}
	\label{fig:freqGa200}
\end{figure}

In the following sections, we will discuss how the effect of the COM offset varies with the density ratio. In doing so, we focus on the representative case of $\Gal = 200$ and $I^* = 1$. We first consider the frequency of the path oscillations ($f$), as this parameter also corresponds to the frequency at which the rotational dynamics are forced. In figure \ref{fig:freqGa200}({\it a\/}), we plot the data in the form of the Strouhal number
\begin{equation}
    \Str = \dfrac{fD}{V_b}.
    \label{eq:Str}
\end{equation}
as a function of $\mt$ and $\Gamma$. The marker colour in the figure indicates the exact $Str$ obtained from the simulations as can be read from the legend, the iso-contours and background colours represent a linear interpolation of this data. The transparent white area bordered by the black dashed line indicates the region where $\gamma > \sqrt{0.5}$. This corresponds to the theoretical state where $I_G$, the MOI of the particle around the centre-of-mass, is zero in accordance with the parallel axis theorem ($I_C = I_G + m_p\ell^2 $) as a consequence of keeping $I_C \equiv m_p D^2/8 = const.$ (\ie $I^* = 1$). Furthermore, we also add a line where $\gamma = 1$, \ie when the point $G$ lies on the particle edge. Results within this marked region are therefore not physically viable, yet still satisfy the governing equations.

Considering the zero offset ($\mt = 0$) cases in figure \ref{fig:freqGa200}({\it a\/}) first, we find that $\Str$ varies quite significantly from $\Str = 0.195$ at high density ratios ($\Gamma \geq 0.2$) to $\Str = 0.127$  for $\Gamma \leq 0.1$. This transition appears to be quite sudden, suggesting the existence of critical density ratio as previously observed for rising and settling blunt bodies \citep{namkoong2008,horowitz2010vortex,mathai2017,Auguste2018,will2021:MOI}. The change in the path frequency at the lowest $\Gamma$ is also associated with an increase in the path amplitude as is evident from the trajectories at $\mt = 0.15$ (which resemble those at $\mt = 0$) in figure \ref{fig:freqGa200}({\it b\/}).

Now let us consider the effects of COM offset for varying $\Gamma$ (\ie moving vertically in the figure). Depending on $\Gamma$, three distinct effects of increased offset can be observed. First of all, for $\Gamma \geq 0.9$ (marked by square symbols throughout), we find that increasing COM offset has almost no effect on the oscillation frequency. There is only a slight decrease for $Str$ at extreme offset as can best be seen in the inset of figure \ref{fig:freqGa200}({\it c\/}). The general lack of response to COM offset in this regime can be explained by considering the rotational equation of motion as presented in \eqref{eq:KK_Tstar}. Remember here that the system is similar to a driven damped harmonic oscillator where the pendulum term is analogous to the spring stiffness, the viscous torque is the damping term, and the accelerated reference frame ($\abc \times \pb$) provides the driving. The restoring torque is proportional to $|1-\Gamma|^{-1}$ and therefore goes to infinity for $\Gamma \to 1$. This is not the case for the driving term which scales according to $\Gamma^{-1}$. Therefore, when the body becomes close to neutrally buoyant, the pendulum torque goes to infinity and as a result the forcing can not rotate the body significantly enough to induce any circulation. Therefore, there will be no Magnus force and no rotational-translational feedback loop leading to resonance. Thus, for the cases where $\Gamma$ is close to unity, the oscillation frequency (as well as other output parameters) of the body remain unaffected by the offset.

The second regime is characterised by a sharp transition in particle dynamics where in a narrow range of $\mt$ the dynamics switch between the base state (near identical to $\gamma = 0$) and the resonance state. This is best shown in the inset of figure \ref{fig:freqGa200}({\it c\/}) where we see that at low values of $\mts$ for intermediate density ratios ($0.2 \leq \Gamma \leq 0.8$) $\Str$ stays constant at approximately 0.195 (upper branch). However, as the offset increases there is a sharp jump to the lower branch of $\Str$. The upper branch corresponds to a system state with minimal body rotation and translation, and in the lower branch the vortex shedding latches on to body motion and is affected by the pendulum frequency. The cases $\Gamma = 0.2$ and $0.8$ are edge cases and show characteristics of their neighbouring regimes.

Finally, the third regime is characterised by a gradual transition to the resonance state and is encountered for $\Gamma \leq 0.1$. Here we find that even at zero offset they are already following the lower branch in figure \ref{fig:freqGa200}({\it c\/}). Since the particle is already exhibiting path oscillations and minute rotational oscillations even at zero offset, no critical threshold of offset needs to be exceeded for the coupling to begin occurring. For these cases, even at $\mts$ below  resonance, we already see offset affecting the particle dynamics. The footprint of these three regimes is also evident in the amplitude and spatial path frequency as shown in figure \ref{fig:freqGa200}({\it b\/}). For high $\Gamma$, there is no effect of increasing offset, at intermediate density ratios we see a large difference between different $\mt$, and at low $\Gamma$, we observe large path oscillations even at small/zero offset.

Beside the jump at the onset of resonance, $Str$ also varies significantly beyond the resonance state. The isocontours of $\Str$ in this region approximately follow the lines of constant $\mts$ (white dashed lines) and in particular the minimum of $\Str$ coincides roughly with $\mts = 0.08$.
The correlation between $\Str$ and $\mts$ is explicitly shown in the inset of figure \ref{fig:freqGa200}({\it c\/}). This plot also highlights the existence of two branches of the system state and the fact the COM offset can trigger a transition between the two, indicated for each $\Gamma$ by the coloured dashed section of the lines. The collapse of the data on these two curves is not trivial and underlines the validity of the Stokes layer argument at the core of the definition of $\mts$. As $\mts$ becomes very large, $\Str$ appears to return to the trend at large $\Gamma$ where higher density ratios have a slightly higher $\Str$. It is further clear that while $\mts$ is the relevant parameter to describe the behaviour after the transition from the low $Str$ to the high $Str$ state, the transition itself does not coincide with $\mts = \textrm{const.}$, but occurs at lower values of $\mts$ for lower $\Gamma$. The case with the lowest density of $\Gamma = 0.001$ spans only a tiny range in terms of $\mts$ even for the largest offsets. This could explain why there is no noticeable variation in the particle behaviour for this density ratio even at large offsets. However, since the driving term is proportional to $1/\Gamma$, it will likely dominate the pendulum torque, which does not diverge for small $\Gamma$.

As mentioned at the beginning of this section, the frequency of the path-oscillations is important for the driving of the rotational dynamics through equation \eqref{eq:KKT_rotation}. As with any harmonic oscillator the parameter of prime importance is the ratio of the driving to natural frequency of the system $f/\tilde{f}_p$, which we show in the main panel of figure \ref{fig:freqGa200}({\it c\/}) as a function of the timescale ratio $\mts$. Curves of constant $\Str$ corresponding to the two different states are indicated by the black dashed lines. Importantly, we find that $f/\tilde{f}_p = 1$ occurs around $\mts = 0.11$, which corresponds to the bold white dashed line in figure \ref{fig:freqGa200}({\it a\/}). We further see in figure \ref{fig:freqGa200}({\it c\/}) that the path oscillation frequency of the body appears to be drawn towards $f_p$ as it begins to deviate from $\Str = 0.127$ to meet $f/\tilde{f}_p = 1$, consistent with the so called lock-in phenomenon \citep{bishop1964lift,bearman1982experimental}. The region of (approximate) frequency lock-in, ranging from $0.09 \leq \mts \leq 0.12$ is indicated by a grey shaded area in the figure background throughout this work. Finally, we included the results for rising and settling spheres with COM offset from the work by \citet{will2021:COM} as black circles in figure \ref{fig:freqGa200}({\it c\/}). The good agreement with the present results suggests that the underlying physics of the resonance mechanism are indeed the same and that results and trends presented here are also relevant for spherical bodies in a 3D flow environment.

\subsection{On the transition to resonance}\label{sec:Special_cases}
\begin{figure}
	\centerline{\includegraphics[width=1.0\textwidth]{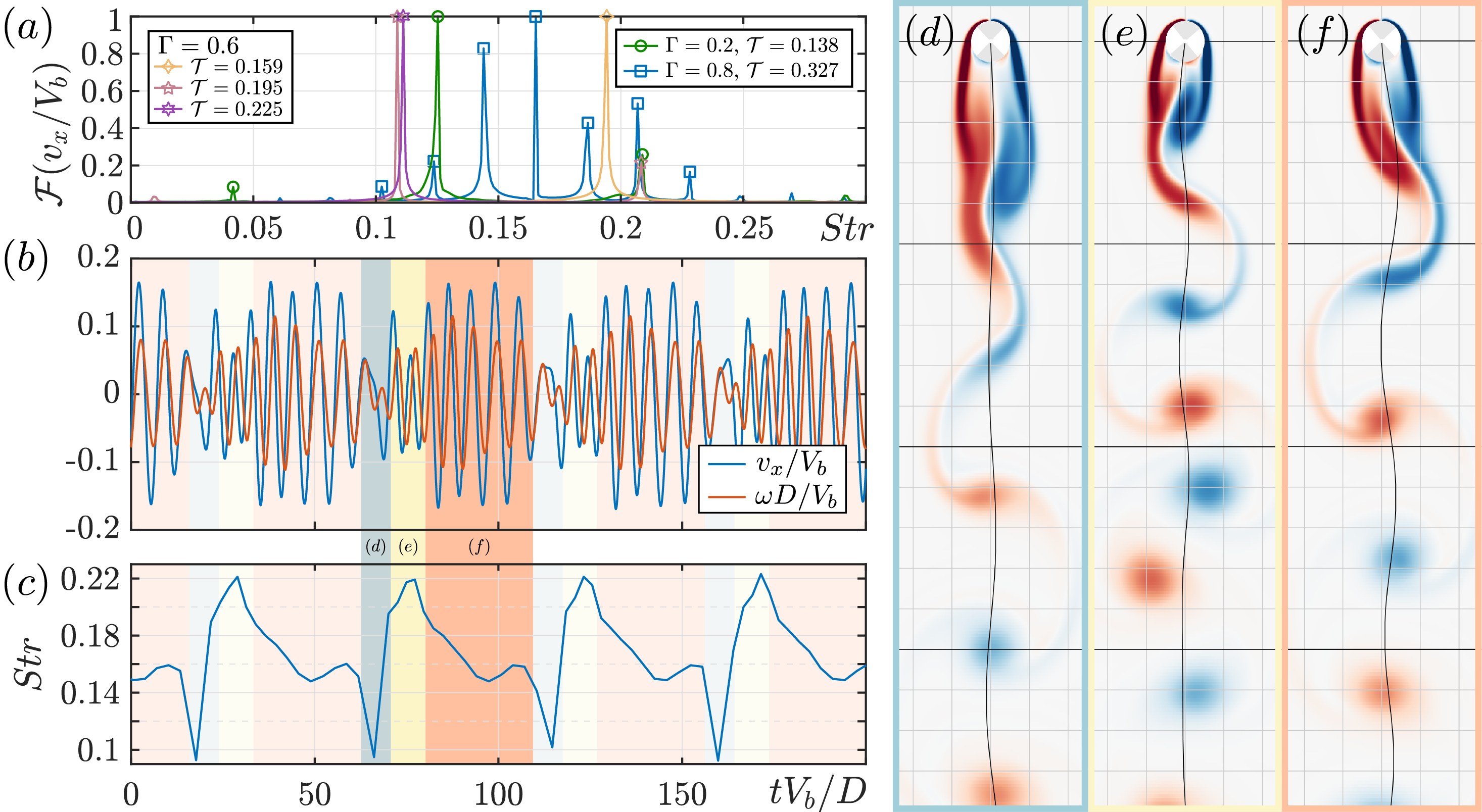}}
	\caption{ ({\it a\/}) Single sided amplitude spectrum ($\mathcal{F}$) based on the particle horizontal particle velocity $(v_x/V_b)$ normalised by the maximum amplitude. ({\it b\/}) Dimensionless horizontal velocity and rotation rate ($\textrm{rad.}/s$) versus time for $\Gamma = 0.8$ and $\mt = 0.327$. We see that the dynamics exhibit a cyclical behaviour on a timescale much greater than that of the vortex-shedding dynamics. This behaviour is split into three parts as indicated by the colours in the background of the figure. ({\it c\/}) Instantaneous Strouhal number as a function of time for $\Gamma = 0.8$ and $\mt = 0.327$, calculated based on the peak-to-peak times of $|v_x/V_b|$. ({\it d--f\/}) Vortex shedding and path oscillations correlating to the modes in ({\it b,\,c\/}). ({\it d\/}) Very low frequency oscillations of minimal amplitude, attached wake is very large. ({\it e\/}) The buildup vorticity is rapidly shed in the wake at a high frequency, resulting in small amplitude high frequency path oscillations. ({\it f\/}) Slower periodic vortex shedding with larger amplitude path oscillations, the attached vorticity slowly grows throughout this phase until the cycle begins anew. 
	}
	\label{fig:Case99}
\end{figure}
In this section, we will discuss the transition to resonance in the intermediate $\Gamma$ regime, i.e. for $0.2 \leq \Gamma \leq 0.8$ in more detail. In the range $0.3 \leq \Gamma \leq 0.7$, the transition from the high $\Str$ number mode to the low $\Str$ one occurs within a narrow band of $\mt$. This is also visible from figure \ref{fig:Case99}({\it a\/}), where the power spectra normalised by the maximum amplitude ($\mathcal{F}$) of $v_x/V_b$ are shown for $\mt = 0.159$, $0.195$, and $0.225$  in the vicinity of the transition point at $\Gamma =0.6$ (yellow diamonds, pink pentagons and purple hexagons respectively). For both, $\mt = 0.159$ and $\mt = 0.225$, the spectra feature singular peaks only at $\Str \approx 0.1$ and $\Str \approx 0.2$, respectively. The former peak also dominates for the intermediate case $\mt = 0.195$, however, a weaker secondary peak at $\Str \approx 0.21$ is also seen to emerge at this offset value. Similar trends can be observed across the range $0.3 \leq \Gamma \leq 0.7$ with varying ratios of relative peak height, suggesting that the transition between modes happens in a narrow band of $\mt$, but is not entirely discrete. 

In \S\ref{sec:freq_of_oscillation}, we mentioned that $\Gamma =0.2$ and $0.8$ were on the edges of the $\Gamma$-range for which a sharp transition to the resonance regime was encountered. We will investigate these cases in more detail here. For $\Gamma = 0.2$, the range of $\mt$ where multiple modes are observed widens significantly as compare to $0.3 \leq \Gamma \leq 0.7$. We observe multiple peaks in the spectra for $0.138 \leq \mt \leq 0.225$ as exemplified by the case shown in figure \ref{fig:Case99}({\it a\/}) (green circles). This extended range is most likely due to the intrinsic rotational and transitional oscillations present at $\gamma = 0$ for $\Gamma$ close to 0 and is similar to the cases of $\Gamma \leq 0.1$. However, looking at figure \ref{fig:freqGa200}({\it c\/}), we still observe that the cases at $\Gamma = 0.2$ and $\gamma \approx 0$ follow the upper branch in terms of $f/f_p$ and \Str, making the behaviour transitional between the $\Gamma$-regimes.

At $\Gamma = 0.8$, we find only a single case ($\mt = 0.327$) for which the system state jumps to the lower branch in figure \ref{fig:freqGa200}({\it c\/}). In figure \ref{fig:Case99}({\it a\/}) (blue squares), we observe that this jump is accompanied by a wide range of frequency peaks. The occurrence of multiple peaks originates from a very unique frequency and amplitude-modulation cycle in the fluid-structure interaction, the signature of which is shown in terms of the time-evolution of $v_x/V_b$ and $\omega D/V_b$ in figure \ref{fig:Case99}({\it b\/}). For both quantities, a modulation of the amplitude but also of the frequency is evident at timescales much larger than that of the path oscillations. This behaviour stands in stark contrast to the rest of the cases which exhibit very regular periodic motion. Specifically, the parameter combinations that show this characteristic behaviour are: $\Gamma = 0.8$ with $\mt = 0.327$ and $\mt = 0.365$, and $\Gamma = 0.7$ with $\mt = 0.411$. A video showing this behaviour along with the vorticity in the wake can be found in \textcolor{black}{supplementary video 2}.

To quantify the frequency variations, we plot the instantaneous $\Str$ in figure \ref{fig:Case99}({\it c\/}) based on determining the distance between the maxima and minima in the signal $v_x/V_b$. Instances where the frequency is relatively low are indicated by the blue regions in figure \ref{fig:Case99}({\it b,c\/}). The corresponding particle kinematics, showing a marked reduction in the lateral amplitude, and the associated vortical structures for the low $\Str$ mode are illustrated in figure \ref{fig:Case99}({\it d\/}). In this case, the cylinder rises almost vertically and the length of the attached wake is at its maximum extent. After this, in the transitional period marked in yellow, the path amplitude remains low but the frequency of the oscillations increases. Figure \ref{fig:Case99}({\it e\/}) shows that this is linked to the rapid shedding of the build-up attached wake, quickly reducing its length. This state is similar to the dynamics observed at higher $\Gamma$. Finally, in the period indicated by red shading, the lateral amplitude is relatively large and the oscillation frequency is intermediate. In the corresponding figure \ref{fig:Case99}({\it f\/}), it is seen that over the course of a number oscillation periods the attached wake slowly grows again until this cycle repeats. Due to the large amplitude and longest duration of this phase, the red region manifests as the strongest peak in the Fourier spectrum and thus the result for $Str$. This behaviour is characteristic for the cases near $\Gamma = 0.8$ and $\mt = 0.327$ at this Galileo number and is indicative of the density regime transitions, where the dynamics exhibit signs of both regimes. These observations highlight that in the transition range, multiple states can coexist.

\subsection{Drag coefficient and Magnus force}
\label{sec:drag}
\begin{figure}
	\centerline{\includegraphics[width=1.0\textwidth]{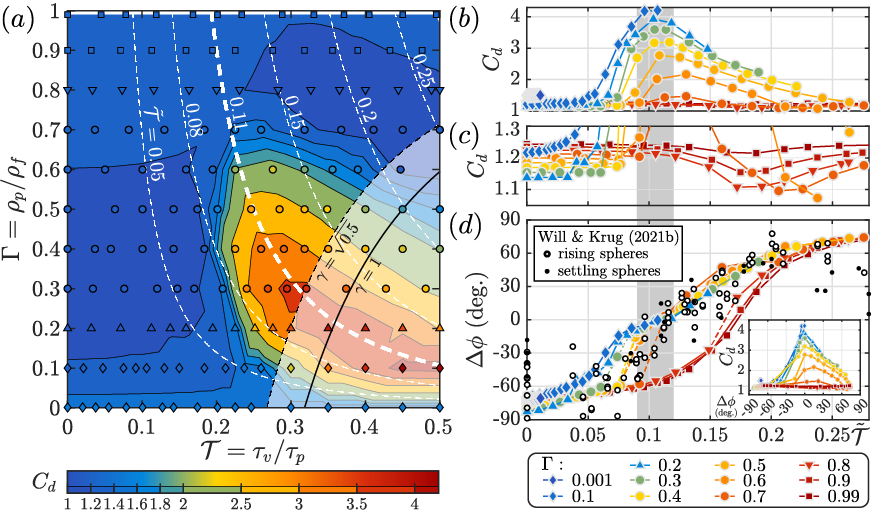}}
	\caption{({\it a\/}) Particle drag coefficient as a function of the particle-to-fluid density ratio $\Gamma$ and the timescale ratio $\mt$ for rising particles at $Ga = 200$ and $I^* = 1$. ({\it b\/}) Drag coefficient plotted explicitly versus $\mts$, the timescale ratio including fluid inertia effects. Capturing the maximum drag trend reasonably well. ({\it c\/}) Zoomed in version of ({\it b\/}) showing the slight reduction of drag beyond the resonance peak.  ({\it d\/}) . Phase lag $\Delta \phi$ between the horizontal Magnus force and the horizontal component of instantaneous acceleration versus $\mts$ alongside the experimental results for spheres \citep{will2021:COM}. The inset shows the correlation between $C_d$ and $\Delta\phi$.
 }
	\label{fig:CdGa200}
\end{figure}

In this section, we will concern ourselves with the mean vertical velocity, \ie the terminal rising/settling velocity, which is of particular practical relevance. We define the drag coefficient $C_d$ obtained from the time-averaged vertical force balance between the buoyancy and the drag force, given that the particle has reached terminal velocity. For a two-dimensional cylinder, this results in
\begin{equation}
    C_d = \dfrac{\pi |1-\Gamma|gD}{2\langle v_y \rangle^2}.
    \label{eq:cd}
\end{equation}
Note that in this definition $C_d$ solely reflects variations in the vertical velocity of the particle. When plotted as a function of $\Gamma$ and $\mt$ (figure \ref{fig:CdGa200}({\it a\/})), the drag coefficient exhibits considerable variations almost up to a factor of 4 across the parameter space that are predominantly induced by COM effects. At $\mt = 0$, $C_d$ is found to be lowest for $\Gamma = 0.2$. Moving to the right in the figure, i.e. towards increasing $\gamma$, the previously (\S\ref{sec:freq_of_oscillation}) defined $\Gamma$ regimes can again be noted. For $\Gamma \geq 0.8$, no increase in $C_d$ is observed as a consequence of the COM offset. However, for $\Gamma \leq 0.7$ the resonance behaviour manifests itself in a strong increase in $C_d$.
These trends are explicitly plotted in terms of $\mts$ in figure \ref{fig:CdGa200}\,({\it b,\,c\/}). For $\Gamma \leq 0.7$, the resonance behaviour reaches a maximum for $\mts \approx 0.11$. This is indicated in figure \ref{fig:CdGa200}({\it a\/}) by the bold white dashed line, and even more evident from the location of the peak in $C_d$ in figure \ref{fig:CdGa200}({\it b\/}). The value of $\mts = 0.11$ corresponds to $f/f^*_p = 1$ in figure \ref{fig:freqGa200}({\it c\/}), \ie where the driving frequency $f$ and the pendulum frequency are identical. The magnitude of the peak drag monotonically increases with $\Gamma$. Beyond $\mts = 0.11$,  $C_d$ gradually decreases again in all resonance cases.
Finally, figure \ref{fig:CdGa200}({\it c\/}) also shows that for all cases at large offsets, even those that did not exhibit resonance (i.e. $\Gamma \geq 0.8$), the drag decreases slightly. It appears that the magnitude of the decrease is inversely correlated with the mass density, resulting in larger reduction for lighter particles (figure \ref{fig:CdGa200}({\it c\/})). This phenomenon does not appear to occur at fixed values of $\mt$ or $\mts$. It is noteworthy, though, that a similar drag reduction was encountered around similar values of $\mts$ for settling and rising spheres with COM offset \citep{will2021:COM}.

In previous work by \citet{will2021:COM}, the connection was made between the drag increase and the maximum in the enhancement of horizontal particle acceleration ($a_x$) through the rotation induced Magnus lift force ($F_{m,\, x} \sim -\omega_z v_y$). 
Besides $F_{m,\, x}$, lateral accelerations can also be driven by pressure fluctuations induced by the vortex shedding. To study the enhancement of the path oscillations via the Magnus force, we consider the phase lag ($\Delta \phi$) between $F_{m,\, x}$ and $a_x$. Examples of time series with different phase lags for three values of $\mt$ are shown in figure \ref{fig:probDef}({\it c\/}). When $F_{m,\, x}$ and $a_x$ are in phase ($\Delta \phi = 0$), the enhancement of the horizontal particle motion by the Magnus force is maximum. The connection between $\Delta \phi$ and $C_d$ is established for the current data set in the inset of figure \ref{fig:CdGa200}({\it d\/}).
In the main panel of figure \ref{fig:CdGa200}({\it d\/}), the phase lag is plotted explicitly vs. $\mts$. Again there is excellent collapse of the data onto two branches, representing the oscillating and non-oscillating states, identical to those encountered for $Str$ in \S\ref{sec:freq_of_oscillation}. We further see that  $\Delta \phi = 0^\circ$ around $\mts = 0.11$ for all cases where resonance is present, whereas acceleration and Magnus force are significantly out of phase ($\Delta \phi \approx  -60^\circ$) in the same range on the lower branch. This point is emphasised by the inset of figure \ref{fig:CdGa200}({\it d\/}), where the peaks in $C_d$ are seen to align with $\Delta \phi = 0$. The good agreement with the experimental sphere data of \citet{will2021:COM}, included in figure \ref{fig:CdGa200}({\it d\/}), again underlining the fact that  the resonance phenomenon in 2D is indeed comparable to that in 3D.

\subsection{Drag correlations}
\label{sec:CdCorr}
 \begin{figure}
	\centerline{\includegraphics[width=1.0\textwidth]{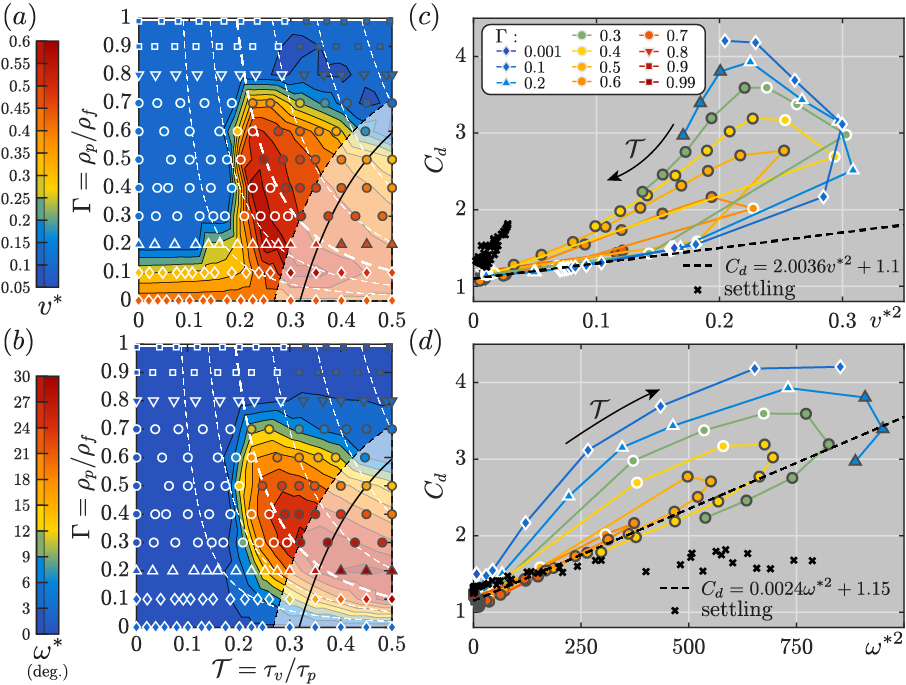}}
	\caption{ ({\it a\/}) Normalised velocity fluctuations $v^* = \sqrt{\langle \vbc'^2 \rangle_\rms} /V_b$ and ({\it b\/}) normalised rotational fluctuations  $\omega^* = \langle \omega \rangle_\rms D/V_b$ (in degrees) as a function of $\Gamma$ and $\mt$. Correlations between the particle drag coefficient and $v^{*2}$ ({\it c\/}) and $\omega^{*2}$ ({\it d\/}). The density ratio ($\Gamma$) in represented by the line colour. The marker infill, white of grey represents $\Delta \phi < 0$ or $\Delta \phi \geq 0$, respectively.}
	\label{fig:Cd_correlations_V2}
\end{figure} 
The question to what extent the drag of freely rising and settling bodies is correlated to path-oscillations and/or particle rotations is subject to an ongoing discussion.
It is indisputable that the presence of horizontal oscillations affects the overall drag coefficient \citep{Horowitz:2010}. However, the presence of rotations was also clearly found to play a prominent role \citep{namkoong2008,Auguste2018,Mathai2018}. In the work on spheres with COM offset by \citet{will2021:COM}, it was shown that the drag correlated better with the mean rotation rate than with the amplitude of the path oscillations or the horizontal velocity fluctuations for cases at or beyond resonance. On the other hand, for zero offset, the drag appeared to correlate equally well with both in the three-dimensional (3D) chaotic regime when varying the MOI of spheres \citep{will2021:MOI} and both of these quantities did not result in an adequate prediction of drag for the spiralling regime.

To add to this, we investigate how variations in $C_d$ relate to the presence and strength of rotational and translational fluctuations in the present data. To this end, we define the dimensionless fluctuating velocity $v^* = \sqrt{\langle \vbc'^2 \rangle_\rms} /V_b$, where $\vbc' = \boldsymbol{v}_C - \langle \vbc \rangle $, presented in figure \ref{fig:Cd_correlations_V2}({\it a\/}), and dimensionless root-mean-squared rotation rate $\omega^* = \langle \omega \rangle_{\textrm{rms}} D/ V_b$, shown in figure \ref{fig:Cd_correlations_V2}({\it b\/}), for all rising cases at $\Gal = 200$. Since particle velocity fluctuations are dominated by their horizontal component, they are qualitatively similar to the amplitude of the path oscillations $\hat{A}/D$. 
A striking difference between the distributions of $v^*$ and $\omega^*$ concerns the region at low density ratios ($\Gamma <0.2$)  and small offsets ($\mt< 0.2$), where significant velocity fluctuations, and hence path-oscillations, are present in the almost complete absence of body rotation (a feature that is different from the findings of \citet{mathai2017} as discussed in Appendix \ref{app:zhu_mathai}). 
In figure \ref{fig:Cd_correlations_V2}({\it c,\,d\/}), we plot $C_d$ vs. $v^*$ and $\omega^*$, respectively. In all panels of figure \ref{fig:Cd_correlations_V2}, the marker edge colour is white if $\Delta\phi<0$ and black if $\Delta\phi\geq 0$. 
For a subset of the markers labelled with a white border,  an approximately linear scaling of $C_d$ with $v^{*2}$ can be observed in figure \ref{fig:Cd_correlations_V2}({\it c\/})). This linear range corresponds to the region of small offsets and low density ratios featuring path oscillations but importantly no rotations. These cases adhere to the scaling $C_d(v^*) = 2.0036 v^{*2} + 1.1$ as indicate by the dashed black line in figure \ref{fig:Cd_correlations_V2}({\it c\/}). However, once rotation begins to become significant, we find that it dominates the drag behaviour. This is demonstrated in figure \ref{fig:Cd_correlations_V2}({\it d\/}), from which it is clear that for the markers with black borders $C_d$ is approximately proportional to $\omega^{*2}$. For cases beyond resonance, results are reasonably well represented by the fit $C_d(\omega^{*2}) = 0.0024\omega^{*2}+1.15$. We find similar quadratic relationships for the higher Galileo number cases examined in this work, but not in the case of settling particles.

\subsection{Settling particles}
\label{sec:settling}

\begin{figure}
	\centerline{\includegraphics[width=1\textwidth]{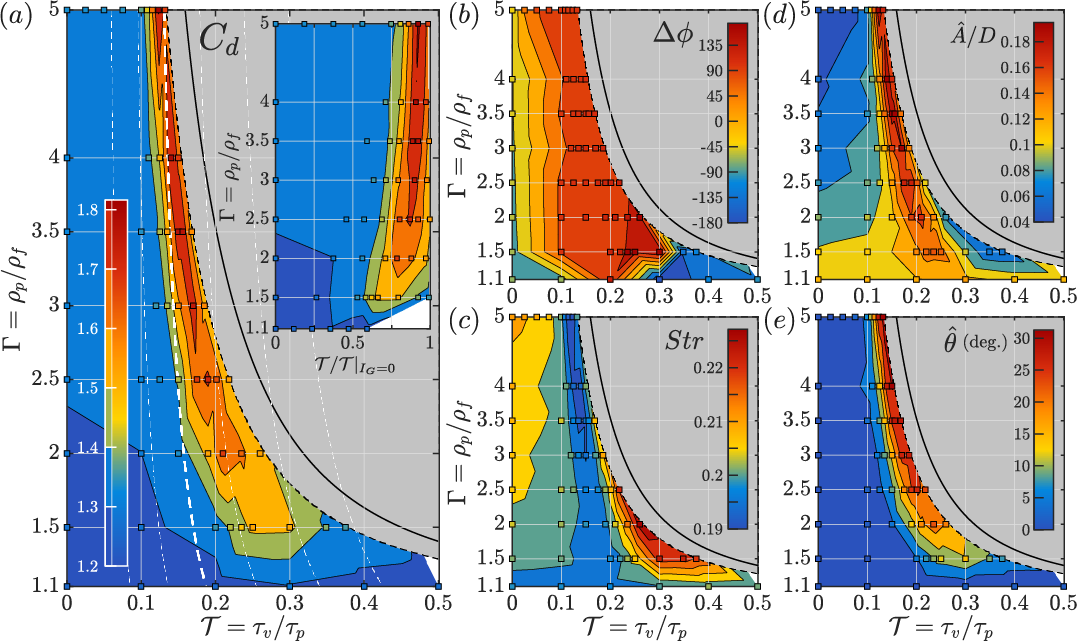}}
	\caption{Results for settling ($\Gamma>1$) 2D cylinders at $\Gal =200$ and $I^* =1$. With ({\it a\/}) showing the  drag coefficient, ({\it b\/}) the Strouhal number, ({\it c\/}) the Phase angle between Magnus force and particle horizontal acceleration, ({\it d\/}) the path amplitude and ({\it e\/}) the rotational amplitude. White lines in (a) represent isocontours of $\mts$.}
	\label{fig:Settling}
\end{figure}

Up until this point, the focus was exclusively on light (rising) 2D cylinders. For heavy particles, it was already demonstrated in \citet{will2021:COM} that the feedback between the Magnus lift force and particle acceleration becomes negative, effectively suppressing the resonance mode. This implies that the strong coupling between rotation and translation and the associated drag increase are absent, but not necessarily that the COM offset has no effect at $\Gamma >1$. To explore this, the density ratio range from $\Gamma = 1.1$ up to 5 was studied with $\mt$ ranging from 0 to the contour $I_G = 0$ at $\Gal = 200$, 500 and 700 and $I^* = 1$. We present only the results for $\Gal = 200$ in figure \ref{fig:Settling}({\it a-e\/}) since the trends for the higher Galileo numbers are similar.

 Figure \ref{fig:Settling}({\it a\/}) confirms that the drag coefficient does vary as a function of $\mt$ also for settling particles. Yet, the magnitude of the increase in $C_d$ (from around 1.2 to 1.8) is much smaller compared to that observed for rising bodies (from around 1.1 up to 4). Furthermore, the drag increase is more pronounced at larger $\Gamma$ and contrary to rising cylinders, the contours of constant $C_d$ do not align well with isocontours of either $\mt$ or $\mts$ (dashed white lines), suggesting that the mechanism of drag increase here is not resonance related. 
 This is further evidenced by figure \ref{fig:Settling}({\it b\/}), where for the the phase lag $\Delta\phi$ is shown for the same cases as in figure \ref{fig:CdGa200}({\it d\/}). Unlike for rising particles, there is no monotonic increasing trend between offset and $\Delta\phi$ and no regime where $\Delta\phi = 0$ can be identified. In fact, the phase lag is strongly positive (between $90^\circ$ and $135^\circ$) in the regions of elevated $C_d$, which implies that $F_m$ (at least in part) counteracts $a_x$.
 The drag behaviour for settling particles rather seems correlated with a reduction of the rotational inertia around point $G$, as the latter tends to zero (black dashed line) for increasing offset due to the fact we maintain $I^* = 1$. In the inset of figure \ref{fig:Settling}({\it a\/}) we show this explicitly by rescaling the horizontal axis according to $\mt/\mt|_{I_G=0}$. Doing so reveals that the drag is maximum for low, but non-zero, rotational inertia (around $\mt/\mt|_{I_G=0}=0.9$). 
 Similarly to rising cylinders, the increase in drag coincides with a decrease in $Str$ (see figure \ref{fig:Settling}({\it c\/})) although this effect is much smaller here as compared to the resonance mode encountered for $\Gamma <1$. Consistent with the trends established for light cylinders at zero offset, the lateral (figure \ref{fig:Settling}({\it d\/})) and rotational (figure \ref{fig:Settling}({\it e\/})) amplitudes are also elevated in this parameter region. This behaviour can also be observed in supplementary video 3 showing six values of $\mt$ for $\Gamma = 2.5$. While the magnitude of all the path-oscillations remains significantly lower as those encountered in the resonance regime for $\Gamma <1$, surprisingly the rotational amplitudes are on a similar level.

The relation between drag and path/rotational oscillations for settling cylinders, shown in figure \ref{fig:Cd_correlations_V2}({\it c,d\/}), is qualitatively similar to that discussed in \S\ref{sec:CdCorr} for rising bodies. However, the exact scaling of $C_d$ with $\omega^{*2}$ is not exactly identical. Furthermore, due to the absence of rotational-translational coupling, the observed increase in body rotation is not reflected in a similar increase in the path-oscillations. We suspect that for settling the increase in drag is primarily resulting from the rotational motion given the minute increase in the translational dynamics in this case. Finally, we observe that the magnitudes of $C_d$, $\hat{A}/D$, and $\hat{\theta} $ all decrease towards $\Gamma = 1$. This behaviour is consistent with the previous results for $\Gamma <1$ and the explanation provided in \S\ref{sec:freq_of_oscillation}, namely the divergence of the pendulum term.

\section{Effects of varying moment of inertia}
\label{sec:moi}
\begin{figure}
	\centerline{\includegraphics[width=1.0\textwidth]{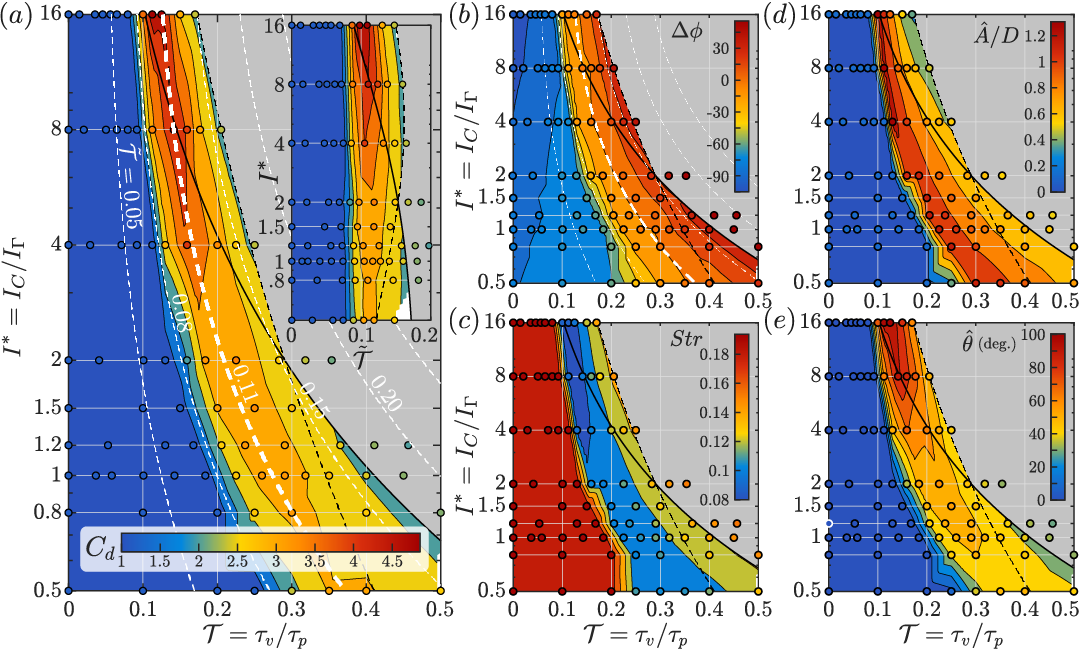}}
	\caption{Investigation on the effect of the dimensionless moment of inertia $I^*$ in combination with the timescale ratio ($\mt$) for $Ga = 200$ and $\Gamma = 0.4$ on the drag coefficient ({\it a\/}), Strouhal number ({\it b\/}), phase lag ({\it c\/}), translational amplitude ({\it d\/}), and the rotational amplitude ({\it e\/}). In these figures the solid and dashed black lines, respectively, represent contours along which $\gamma = 1$ and $I_G = 0$. In ({\it a\/}) the inset shows the same data as the main panel, however is plotted against the modified timescale ratio $\mts$ to include effects of the rotational added mass due to the Stokes layer.}
	\label{fig:moi}
\end{figure}

In this section, we will explore the effects variations in the particle MOI around $G$, which thus far has been kept constant. Since particle rotation proved to be a critical aspect in the preceding analysis, it is anticipated that the variations of the MOI will also affect the overall dynamics. We investigate this by varying the dimensionless MOI around the geometric centre ($I^*$) in the range $I^*\in[0.5,16]$ for the cases of $\Gamma= 0.4$, $Ga=200$ and $\mt\in[0,0.5]$. The corresponding results for $C_d$, $\Delta \phi$, $\Str$, $\hat{A}/D$, and $\hat{\theta} $ are presented in figure \ref{fig:moi}. In these figures, physically feasible boundaries are represented by the solid black line, which indicates the line where $\gamma = 1$ and the dashed black line indicates where $I_G = 0$. The grey shaded region marks parameter combinations beyond both these two criteria. This region is probed less extensively and therefore no linear interpolation of the data is provided there.

The results for $C_d$ as a function of $I^*$ and $\mt$ in figure \ref{fig:moi}({\it a\/}) clearly underline the need to include the fluid inertia in the analysis of the problem. This is obvious from the fact that the $\mt$ values at the maximum in $C_d$ show significant variation as a function of $I^*$, while inclusion of the fluid inertia in the definition of $\mts$ resolves this dependence.
The latter can be seen from the white dashed lines in figure \ref{fig:moi}({\it a\/}), but is even more evident from the inset, where the same data is plotted directly vs. $\mts$; the maxima in $C_d$ collapse onto $\mts = 0.11$. Besides the dependence on $\mts$, our data further show that higher values of $I^*$ lead to an increased peak drag coefficient at resonance.

In figure \ref{fig:moi}({\it b\/}), $\Delta\phi$ is shown for the same data set. Identically to the results described in \S\ref{sec:drag}, we find that peak drag occurs for $\Delta\phi = 0$ when the system is in resonance. The phase data is the best way to asses the validity of the inclusion of fluid inertia as shown in figure \ref{fig:moi}({\it b\/}), the isocontours of $\mts$ almost exactly match the interpolated $\Delta\phi$ data proving the efficacy of this model. Note that this match is obtained with no additional fitting such that this validates also our choice for the value of $I_a^*$, that was obtained based on $Ga$ trends in \S\ref{sec:fluidInertia}. 

In figure \ref{fig:moi}\,{\it c--e\/}, we present corresponding results for the Strouhal number and for the translational  and rotational amplitudes, respectively. Consistent with the observations for $C_d$ and $\Delta\phi$, isocontours of all these quantities also line up with lines for which $\mts = const$. Also in line with the $C_d$ results, the resonance induced changes become stronger with increasing $I^*$ in all quantities considered. 
This behaviour can be understood by considering the systems as a driven and damped harmonic oscillator noting that when the rotational inertia of the system increases, the damping ratio decreases resulting in larger rotational amplitudes $\hat{\theta} $. This, in turn, then affects the other parameters ($C_d$, $Str$ and $\hat{A}/D$), for which the known effects of rotation become enhanced.

As an aside, we would like to remark on the zero offset case, $\gamma = \mt = 0$, which was studied in detail in \citet{mathai2017}. Based on their simulations, these authors identified a transition in particle dynamics and vortex shedding mode as a function of $I^*$. We were unable to reproduce such a transition in our simulations. In order to clarify this difference, a direct comparison of these two contradictory results is provided in Appendix \ref{app:zhu_mathai} at matching Galileo number ($Ga =500$) and overlapping range of $I^*$ and $\Gamma$. Finally, we would like to point out that the conclusion based on the present data, i.e. that rotation plays a very marginal role in affecting regime transition in absence of a COM offset, is also consistent with the experimental study on the effect of varying MOI for rising spheres by \citet{will2021:MOI}.

\section{Galileo number effects}
\label{sec:Galileo}
\begin{figure}
	\centerline{\includegraphics[width=1\textwidth]{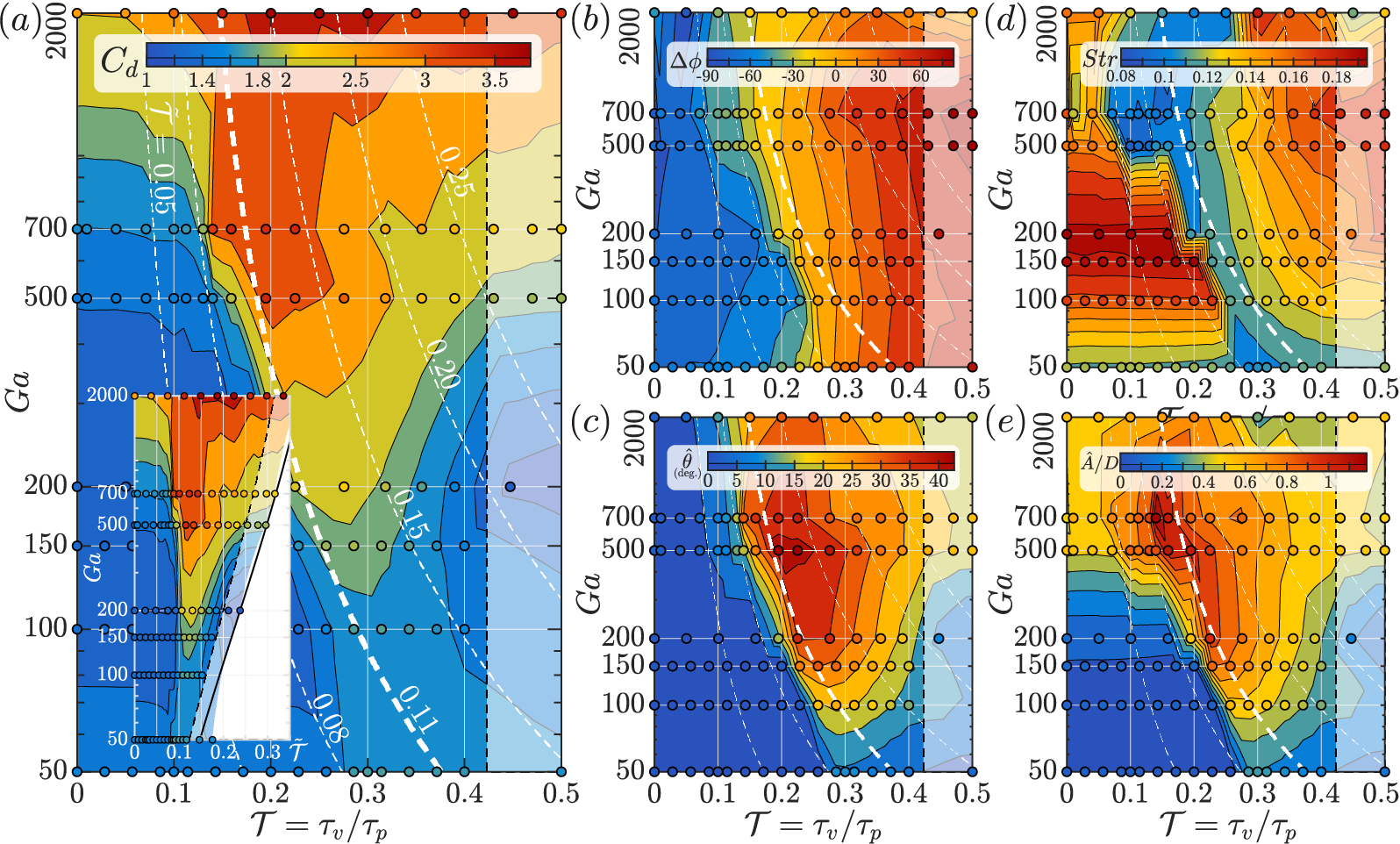}}
	\caption{Exploration of the combined effects of COM offset and Galileo number for fixed density ratio $\Gamma = 0.6$ and dimensionless moment of inertia $I^* = 1$. With ({\it a\/}) showing the drag coefficient $C_d$ and the inset highlighting the scaling in terms of $\mts$, ({\it b\/}) the phase lag $\Delta\phi$, ({\it c\/}) the mean rotational amplitude $\hat{\theta}$, ({\it d\/}) the Strouhal number $\Str$, and ({\it e\/}) the mean path-amplitude $\hat{A}/D$ . The white dashed lines indicate isocontours in $\mts$, the black dashed line indicates $\gamma = \sqrt{0.5}$.} 
	\label{fig:HGC}
\end{figure}

In this section we revisit the Galileo number, previously discussed in \S \ref{sec:fluidInertia}, however here we will take a broader scope and will look beyond the effects of fluid inertia. In this work seven Galileo numbers, ranging from 50  up to 2000, were examined for varying COM offset at fixed $\Gamma = 0.6$, and $I^* = 1$. These results are presented in figure \ref{fig:HGC}. Furthermore, for $\Gal$ = 500 and 700 we also varied $\Gamma$ from 0.001 up to 5 identical to what was previously presented for $\Gal = 200$. The only difference in the results for higher Galileo number settling particles is that for $0.16 < \mt < 0.19$ a significant horizontal drift is encountered $v_d/V_b > 0.1$, resulting in oblique trajectories. Results for the drift velocity for all cases can be found in the supplementary data.

In figure \ref{fig:HGC} ({\it a\/}) the drag coefficient is shown as a function of $\Gal$ and $\mt$. For all cases depicted here, an increase in drag associated with COM offset is observed. The magnitude of this increase in drag is found to become larger with increasing Galileo number. At $\gamma = 0$, a minimum drag ($C_d = 1.2$) occurs for $\Gal = 200$. The increase in $C_d$ towards lower $\Gal$ is related to the viscous dominance in this regime and for higher $\Gal$ the increase in $C_d$ is associated with increasing path and rotational oscillations. Furthermore, the isocontours of $\mts$, including fluid inertia effects, capture the essential features of Galileo number dependence of the $C_d$ variation reasonably well. This is highlighted in the inset of figure \ref{fig:HGC} ({\it a\/}), where we rescale the horizontal axis to $\mts$. This also reveals that the onset of resonance occurs near at constant $\mts$. Additionally, the range of offsets where $C_d$ is affected extends to larger $\mts$ for increasing $\Gal$ in a similar way as decreasing $\Gamma$ or $I^*$ would.

Figure \ref{fig:HGC} ({\it b\/}) shows $\Delta \phi$ for the same parameter range. 
Here, the $\mts = \textrm{const.}$ contours match with constant $\Delta\phi$ predominantly over the range $-45^\circ < \Delta\phi < -15^\circ$ ($0.08<\mts <0.11$). This is a consequence of our choice to base 
the value of $I^*_a$ and more specifically the fitting coefficient $c_1$ on collapsing the rising edge of $\langle \hat{\theta} \rangle$, figure \ref{fig:fluidInertiaEffects}\,({\it c\/}). An alternate choice would have been to fit $c_1$ to align $\Delta \phi =0$ across all $Ga$. Doing so results in a value $c_1$ of approximately $0.5$ (note that this yields a more than 80\% reduction in $I^*_a$ for $\Gal < 2000$), resulting in $\mts$ contours overlapping the $\Delta \phi = 0$ isocontour in figure \ref{fig:HGC} ({\it b\/}). This does not alter the conclusion that added rotational mass is responsible for the observed behaviour. Previously discussed results collapse in a similar way for both $c_1 = 0.5$ or $c_1 = 2.3$, but the differences highlight the fact that the actual value depends on the particle dynamics and kinematics for a given parameter combination of $\mt$, $\Gamma$, $I^*$, and $\Gal$ and is, in fact, not constant in time altogether.

For all Galileo numbers a reduction in the Strouhal is encountered around $\mts = 0.11$, see figure \ref{fig:HGC} ({\it d\/}). This drop is contingent on the presence of rotational oscillations: when rotation is absent the path-oscillation frequency is high and when it is present it is much lower identical to the behaviour observed in \S\ref{sec:freq_of_oscillation}. Based on the results here rotational amplitudes of approximately $5^\circ$ at high Galileo numbers are sufficient to cause a drastic drop in $\Str$.

In figure \ref{fig:HGC}({\it e\/}) the path-amplitude response is depicted. The behaviour here is similar to that of $C_d$ and $\hat{\theta}$ except for the large increase in path-amplitude at high $\Gal$ and small values of $\mt$. We can see that as $\Gal$ increases, the path amplitude at $\gamma = 0$ also grows for $\Gal \geq 500$, whereas the rotational amplitude remains low in the same parameter region (see figure \ref{fig:HGC} ({\it c\/}). This behaviour is akin to that observed for spheres as reported in \citet{Auguste2018,will2021:MOI}, where a $\Gamma$ threshold is encountered demarcating the transition between a vertical and chaotic rise mode, notably in the absence of strong rotation. The $\Gamma$ value of this threshold was shown to increase with $\Gal$ which is what we find here as well since for $\Gal =200$ this transition was encountered for $0.1 < \Gamma < 0.2$, see \S \ref{sec:freq_of_oscillation}. And indeed, for $\Gal \geq 500$ the behaviour of the 2D cylinders is chaotic with large fluctuations in both path and rotational amplitudes without period-to-period regularity. 

\begin{figure}
	\centerline{\includegraphics[width=1.0\textwidth]{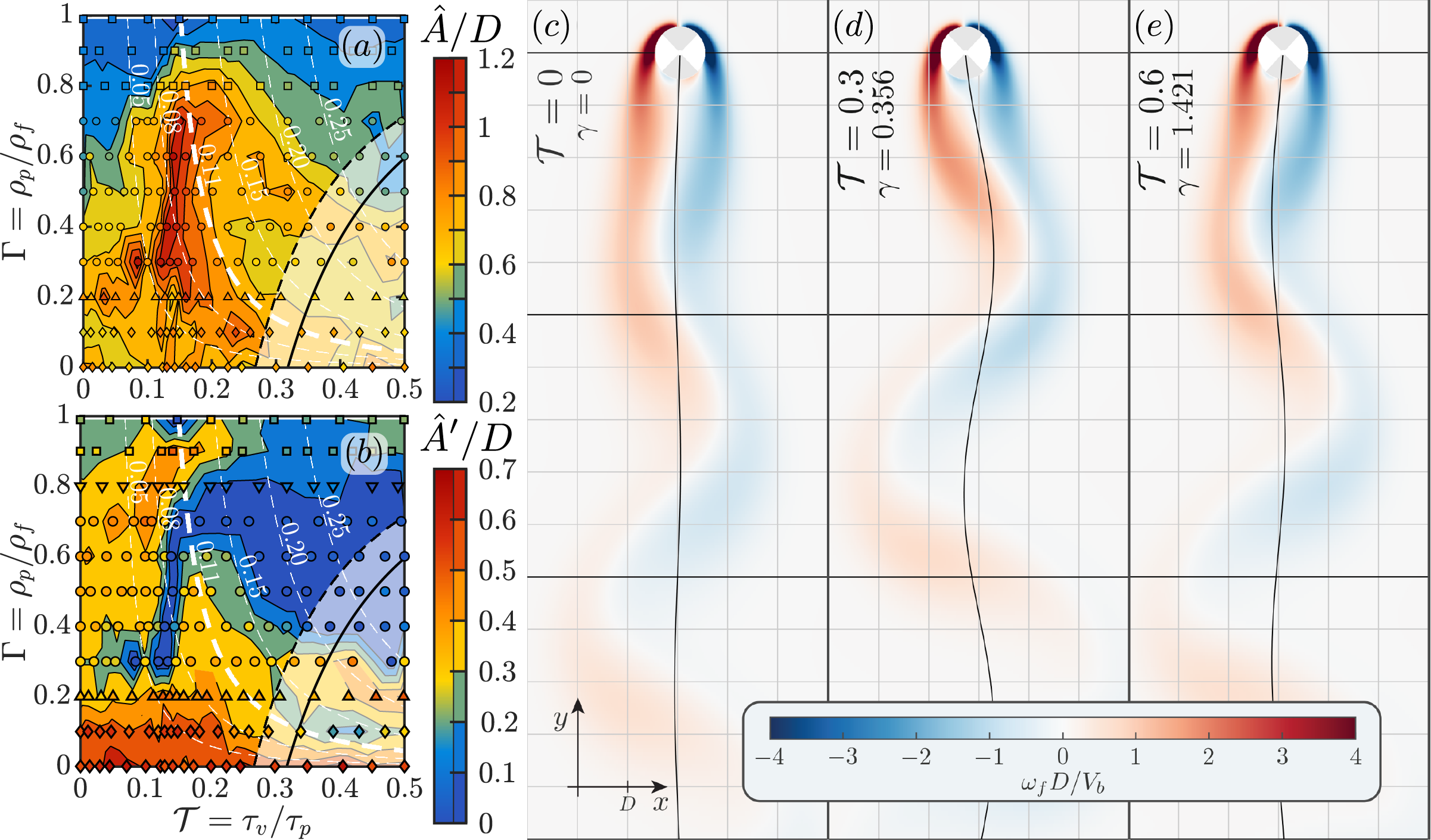}}
	\caption{({\it a\/}) Amplitude of the path oscillations for $\Gal = 700$ and $I^* = 1$ as a function $\mt$ and $\Gamma$. ({\it b\/}) For the same parameter space shows the standard deviation of the peak path-amplitude. ({\it d--e\/}) Trajectories and wake structure for $Ga$ 50, $\Gamma = 0.6$, $I^* = 1$ and $\mt = 0$, 0.3, and 0.6, respectively. The colour gradient in the wake indicates non-dimensional fluid vorticity ($ \omega_f D/ V_b$) as indicated by the colour bar.}
	\label{fig:Ga50}
\end{figure}

To highlight this period-to-period irregularity and to demonstrate the effect of COM offset on these dynamics, we show the path-amplitude ($\hat{A}/D$) for $\Gal = 700$ in figure \ref{fig:Ga50} ({\it a\/}) as well as the standard deviation of $\hat{A}'/D$ of this quantity in figure \ref{fig:Ga50} ({\it b\/}). For these higher Galileo number cases the magnitude of the fluctuations in the amplitude is comparable to the path-amplitude itself. The irregularity is also much higher than at  $\Gal = 200$, for which $\hat{A}'/D \leq 0.1$ even for the transitional cases discussed in \S \ref{sec:Special_cases}. This stresses that behaviour at higher $\Gal$ is in fact chaotic and it is therefore quite remarkable that the mean quantities are reasonably well behaved and in-line with results for lower Galileo numbers. This can in part be explained by a second observation pertaining to figure \ref{fig:Ga50} ({\it b\/}), namely that when the resonance threshold is exceeded, \ie $\mts > 0.11$, the value of $\hat{A}'/D$ drops drastically. The periodicity imposed by the pendulum frequency becomes dominant and stabilises the chaotic motion resulting from the body-fluid interaction. Qualitatively, this chaotic behaviour at low offset and the stabilising effect of large offsets is visible in the \textcolor{black}{supplementary video 4 and supplementary video 5} showing results for $\Gal =  500$ and $700$, respectively.

Finally, we focus on the lowest Galileo number ($\Gal = 50$), where the zero offset case exhibited a vertical (non-oblique) rise mode with superimposed path oscillations. For these cases, the discrete vortex shedding that is characteristic of high $Re$ flow around a blunt body is no longer observed. Instead, an oscillating wake is encountered as depicted for the $\gamma = 0$ case in figure \ref{fig:Ga50}\,({\it c\/}), where the particle path and wake pattern (in terms of fluid vorticity) are visualised. Importantly, however, we find that even the small pressure asymmetry induced by the oscillating wake and the associated small path-oscillations ($\hat{A}/D = 0.06$ at $\gamma =0$) combined with COM offset are enough to trigger resonance behaviour similar to that observed for the $\Gal = 200$ case. In figure \ref{fig:Ga50} ({\it d\/}), a snapshot of the simulations for the resonance case is shown, featuring larger amplitude path oscillations and a wider wake. Similar to the $\Gal=200$ case, increasing the offset beyond resonance again leads to a reduction of the amplitude of the oscillations as shown in figure \ref{fig:Ga50}\,({\it e\/}). Based on this result, we conclude that COM offset will affect the dynamics as long as the base state at $\gamma = 0$ exhibits path-oscillations.

\section{Summary and conclusions}\label{sec:conclusion}
In this work, we have systematically studied Centre-Of-Mass (COM) offset effects for a freely rising or settling cylinder in a quiescent fluid via direct numerical simulations employing the Immersed Boundary Projection Method.
The non-dimensional parameter characterising the COM offset is given by the timescale ratio (\ref{eq:T}) $\mt\equiv \tau_v/\tau_p $, with $\tau_v$ defining the vortex shedding frequency timescale (set by the buoyancy velocity and particle diameter), and $\tau_p$ a timescale originating from the ``pendulum''-like restoring torque resulting from the offset between the centres of mass and buoyancy \citep[\cf][]{will2021:COM}. The main goal of this work has been to confirm that the behaviour of COM offset can be predicted in term of this timescale ratio, which depends on both $\Gamma$ and $I^*$. Additionally, a dependence on the Galileo number is expected but this is not reflected in the definition of $\mt$ by \citet{will2021:COM}. These dependencies could not be adequately confirmed experimentally in previous work due to physical constraints, therefore, a numerical study was desirable since then one can exactly set all of the control parameters. Simplification to 2D, i.e. cylinders, allowed us to examine the 4-dimensional control parameter space which is not feasible for 3D. The thus studied parameter space ranges $0 \leq \mt \leq 0.6$ for COM offset for Galileo numbers ranging from 50 up to 2000, $0.001 \leq\Gamma \leq 5$, and $0.5 \leq I^* \leq 16$.

First of all, we found that for rising particles the dynamics and response to the offset was qualitatively similar to that of spheres; a resonance mode was encountered at a particular offset for which the particle rotation and drag are significantly enhanced. For increasing offset, this effect slowly reduces towards a case where no rotation is present. This behaviour at larger $\Gamma$, constant (or large) $\Gal$, and constant $I^*$ appeared to be well described by $\mt$ (as was the case for \citet{will2021:COM}).  However, for lower $\Gamma$ or $\Gal$ we found that an additional effect was playing a role. This was identified as the contribution of rotational fluid inertia ($I_a$). We modelled this contribution as an annulus surrounding the cylinder, the thickness of which scales according to a boundary layer ($1/\sqrt{\Gal}$, which is identical to $1/\sqrt{\Gal}$ when $C_d$ is constant). This hence introduces a Galileo (or Reynolds) number dependence in the definition of the timescale ratio resulting in (\ref{eq:Tstar}), that was previously not explored. For rising cylinders, for which the resonance phenomenon is present, this modified timescale ratio ($\mts$) was seen to capture the trends in the observed resonance behaviour with resonance occurring at $\mts = 0.11$.

Secondly, we find that body rotation is of crucial importance when considering the dynamics and kinematics of a body moving through a fluid. When altering COM offset, only the rotational equation of motion of the body is affected, and indeed we note a substantial increase in rotation around $\mts = 0.11$. But more importantly, this also affects the frequency of oscillation, path-amplitude, and drag coefficient (terminal velocity). Altering the COM a couple of percent can induce a more than three-fold increase in $C_d$. This increase in drag can be attributed to an increase in both rotational and path oscillations. However, the effect of rotation typically is way more significant and instances exhibiting translational oscillations without rotation featured only relatively small drag increases. While $\mts$ describes the behaviour of the output parameters relative to the offset, the magnitude of the variation (such as the drag increase) still depends on the full parameter combination; for instance the magnitude of the drag increase in resonance still depends on both $\Gamma$ and $I^*$.

Thirdly, we determined that the driving of the rotational dynamics for bodies with COM offset originates from the torque generated by horizontal path-oscillations, i.e. the $\abc \times \pb$-term in (\ref{eq:Tstar}). When switching only this coupling term off in the equations of motion while leaving the pendulum term unchanged, there was almost no difference in the particle behaviour with or without offset for both rising and settling particles. Conversely, the (viscous) fluid torque almost entirely acts as a damping term in the presence of an offset limiting the rotational oscillations. Generally, the magnitude of the viscous torque was found to be too low to drive significant rotations. This applies also at zero offset where the present data did not reproduce the regime transition for varying rotational inertia reported in \citet{mathai2017}. 

Finally, we confirmed that for $\Gamma > 1$, i.e. settling cylinders, the resonance phenomenon is no longer present. As outlined by \citet{will2021:COM}, the feedback between the rotation induced Magnus lift force and the direction of horizontal acceleration becomes negative for heavy particles. Nevertheless, some effects of the offset also exists for $\Gamma > 1$, however, they occur at larger offsets than expected based on the light counterparts and the effect on $C_d$ is significantly smaller (and importantly does not scale according to $\mts$). Instead of the resonance mechanism the explanation for this behaviour appears to be related to the reduction of $I_G$ resulting from us enforcing $I^*=1$. Surprisingly, we also uncover that for both rising and settling no effect of COM occurs when $\Gamma$ is around unity. This can be explained by the fact that the ratio of the pendulum torque over the driving torque ($\Gamma/|1-\Gamma|$), in (\ref{eq:Tstar}) goes to infinity for $\Gamma \to 1$. Thus the driving can not overcome the restoring force of the pendulum and rotations remain too small to engage the feedback mechanism.\\

To summarise, we have given a complete overview of how COM offset depends on the four control parameters governing the system for both rising and settling cylinders in a quiescent fluid. The dynamics and kinematics uncovered here in terms of $\mts$ qualitatively match those uncovered by \citet{will2021:COM} for rising and settling spheres. This suggests that the present findings largely transfer to the behaviour of spherical particles.  Especially, this concerns the behaviour at low Galileo numbers, where also for spheres fluid inertia will become increasingly important which can be accounted for analogously via an added mass term.

\section*{Acknowledgements}
We acknowledge PRACE for awarding us access to MareNostrum at Barcelona Supercomputing Center (BSC), Spain (Project 2020225335). This project has further received funding from the European Research Council (ERC) under the European Union's Horizon 2020 research and innovation programme (grant agreement No. 950111, BU-PACT), as well as from the Netherlands Organisation for Scientific Research (NWO) under VIDI Grant No. 13477 and through the research programme of the Foundation for Fundamental Research on Matter with project number 16DDS001, which is financially supported by the NWO.\\

\noindent \textbf{Declaration of interests.} The authors report no conflicts of interest.

\appendix

\section{Verification and validations of the immersed boundary projection method}\label{app:eomparticle}

\begin{figure}
	\centering
	\includegraphics[width=0.6\linewidth]{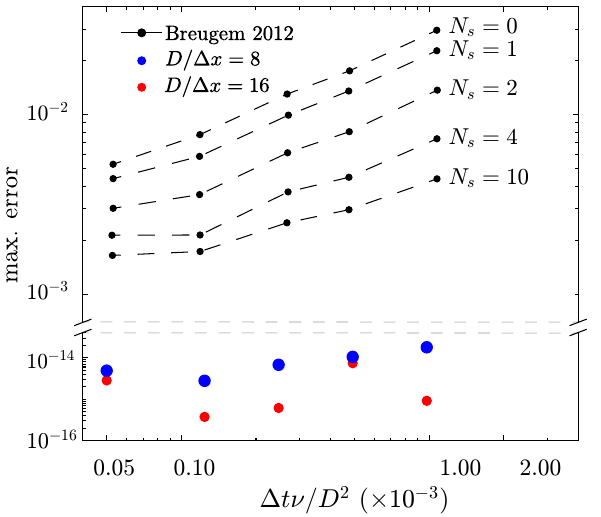}
	\caption{
 	Comparison of the multidirect forcing \citep{breugem2012} versus the immersed boundary projection method. The error of the no-slip condition is given in the maximum norm for various $\Delta t$ and $\Delta x$. The current approach is found to accurately represent the no-slip condition for $\qb^n$, which verifies a correct implementation of the no-slip Lagrangian multiplier.}
	\label{fig:bccheck}
\end{figure}

In order to validate and verify the correct implementation of the immersed boundary projection method, we performed a series of checks and compared our results to those available in the literature.

\subsection{Accuracy verification of the Lagrangian multipliers}
In the following, we present the performance of the immersed boundary projection method (IBPM). First, we test how accurate $\ub^{n+1}$ complies with the Lagrangian multipliers, \ie the newly implemented pressure solver should produce solutions that are divergence-free, and provide a machine accurate no-slip condition at the particle edge (see equation \ref{eq:constraints}). To this point, a freely moving cylinder with $\Gamma=0.001$ is released at $\Gal=10$ in a domain of $16D\times16D$. We present results for two cases with an evenly spaced grid of $128\times128$ and $256\times256$ grid points, yielding 8 to 16 grid points per diameter, respectively. The case is integrated in time until the cylinder reaches terminal velocity, after which statistics are collected.

The maximum divergence, observed via $G^T\qb^{n+1}$, was found to be of $\mathcal{O}(10^{-14})$, which verifies that the velocity field is divergence-free. Next, let $\eb=E\qb^{n+1} - (\vbc^{n+1}-\omb\times\Lcalb)$ define the error vector of the no-slip condition.

The maximum error of the velocity boundary condition  $\max|\eb|$ is presented for various times steps in figure \ref{fig:bccheck}. To put the results in perspective, we compare them to the findings of \cite{breugem2012} who studied the accuracy of the multi-direct forcing for a fixed sphere. Overall, results for $L_\infty$ are found to be of $\mathcal{O}(10^{-14})$ or lower, which confirms that the pressure solver accurately obtains the solution $\phib^{n+1/2}$ such that $\qb^{n+1}$  satisfies the no-slip condition. Achieving a much more accurate no-slip condition up to machine precision is a clear advantage over the multi-direct forcing scheme, but is computationally more demanding.

\subsection{The order of convergence}
In the following we assess the convergence properties of the fluid solver coupled to the particle equations of motion. Here, we select a two-dimensional problem of a cylinder under the influence of gravity in a viscous fluid, which is initially at rest. The domain is square and has a width of $12.288D$. The particle-to-fluid density ratio is set to $\Gamma=1.1$, and the Galileo number to $\Gal=100$. We position the cylinder at the centre of the domain. For each case, we use constant grid spacing and time steps.

\begin{figure}
	\centering
	\includegraphics[width=0.85\linewidth]{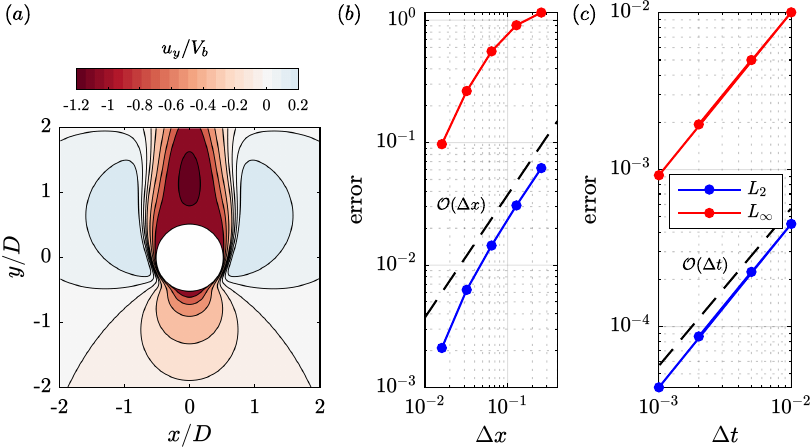}
	\caption{Convergence analysis for the vertical velocity field around a settling cylinder. $(a)$ Snapshot of $v_y/V_b$ at time instance $t=6D/V_b$ for the reference case $(\Delta x= 8\times 10^{-3})$ of the spatial convergence analysis. $(b)$ Spatial convergence.  $(c)$ Temporal convergence with constant $\Delta x=1.6\times 10^{-2}$ for each case, and a reference temporal spacing of $\Delta t = 1\times 10^{-4}$.}
	\label{fig:velocityconvergence}
\end{figure}

\begin{figure}
	\centering
	\includegraphics[width=0.85\linewidth]{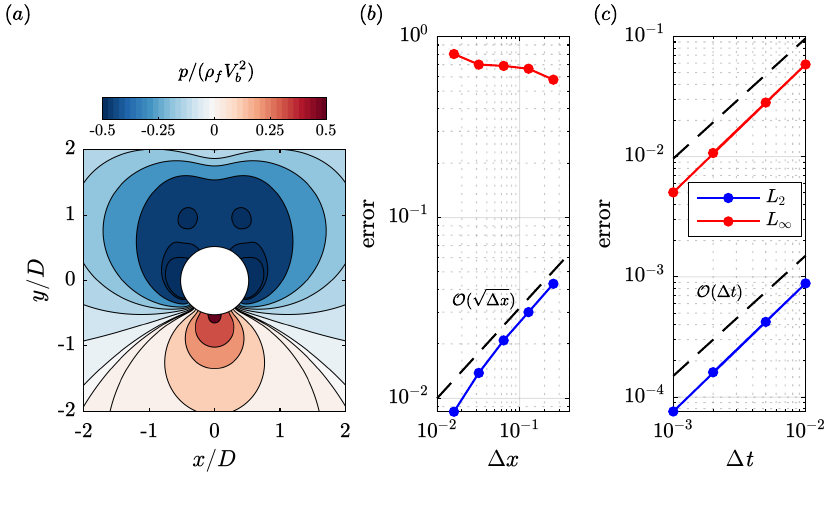}
	\caption{
		 Convergence analysis for the pressure field around a settling cylinder for the same problem as in figure \ref{fig:velocityconvergence}. $(a)$ Snapshot of $v_y/V_b$ at time instance $t=6D/V_b$ for the reference case $(\Delta x= 8\times 10^{-3})$ of the spatial convergence analysis. $(b)$ Spatial convergence.  $(c)$ Temporal convergence with constant $\Delta x=1.6\times 10^{-2}$ for each case, and a reference temporal spacing of $\Delta t = 1\times 10^{-4}$. }
	\label{fig:pressureconvergencecompilation}
\end{figure}

First, we investigate the spatial and temporal convergence of the vertical velocity field. We start with the spatial convergence by fixing the time step to $\Delta t =1.0\times 10^{-4}D/V_b$, and vary the grid spacing $\Delta x \in [0.016, 0.256]$. After a time period of $6D/V_b$ we compare the vertical velocity field $u_y$ for each case to that of the reference case ($\Delta x = 0.008$). We present a snapshot of the latter in figure \ref{fig:velocityconvergence}(a). Note that at this point in time, the horizontal particle velocity is still negligible ($v_x/v_y=\mathcal{O}(1\times 10^{-12}))$.  Let $e_i$ define the error between the vertical velocity field of a simulated case and the reference case, \ie $e_i= [
{u}^{ \textrm{ref}}_{y}(i) - u_{y}(i)]/V_b$, with ${u}^{\textrm{ref}}_{y}(i)$  the reference velocity interpolated onto the coarser grid using a third order interpolation method. Note that ${u}^\textrm{ref}_{y}(i)$ and $u_{y}(i)$ are rewritten from a two-dimensional field to the vector form.
In this study, we shall make use of the $L_2$ and $L_\infty$ norm to report the convergence rate of the implemented IBPM. For this, we define a normalised non-negative vector norm as 
\begin{equation}
	\|\xb\|_p = \left(\dfrac{1}{\Omega_0}\sum_i|x_i|^p \Delta x_i\Delta y_i\right)^{1/p},
\end{equation}
with $p=1,2,\ldots$ an integer, and $\Omega_0$ the corresponding domain surface. For the $L_2$-norm and $L_\infty$-norm it is readily obtained that
\begin{equation}\label{eq:l2_linf_norm}
	\|\xb\|_2 = \sqrt{\xb^T\xb/N}\quad\textrm{and}\quad \|\xb\|_\infty = \max(|x_i|),
\end{equation}
when $\Delta x_i = \Delta y_i = \Delta x$, and $N$ the number of entries in the field.

Figure \ref{fig:velocityconvergence} presents the spatial convergence results. We observe that the $L_2$ error converges at a rate of $\mathcal{O}(\Delta x)$, whereas the  $L_\infty$ norm becomes of $\mathcal{O}(\Delta x)$ for small enough grid spacings only. These results for the convergence are consistent with those reported in \cite{stein2017} for a two-dimensional test problem. Note that \citet{lacis2016} employed $\|\xb\|_2 = \sqrt{\xb^T\xb}/N$ instead of the definition for $\|\xb\|_2$ provided in \eqref{eq:l2_linf_norm}, where the present definition is preferred because it is consistent between continuous and discrete forms. This difference explains the somewhat faster convergence rate reported in their case.

Next, we investigate the temporal convergence for $u_y$, by selecting a grid spacing of $\Delta x = 0.016D $ for each case, and vary the time step  $\Delta t \in[1.0\times 10^{-3},\,1.0\times 10^{-2}]D/V_b$. A reference case is chosen with the same grid spacing and a smaller time step of $\Delta t =1.0\times 10^{-4}D/V_b$. At $t=0.9D/V_b$ the convergence results are assessed. Note that the reference case has reached a velocity of $v_y\approx0.32V_b$ at this point in time.
The results of $L_2$ and $L_\infty$ for the temporal convergence test are depicted in figure \ref{fig:velocityconvergence}$(c)$. Here, we observe first-order convergence for both norms. This result is expected due to the approximation of $A^{-1}\approx B_1$, where the leading term of $B$ is of $\Delta t$ and is in agreement with that of \cite{lacis2016} who observed the same temporal convergence for a freely falling cylinder.

Now, we turn our attention to the spatial and temporal convergence of the pressure field. The datasets for this analysis are the same as those used for the vertical velocity field. We start with the analysis of spatial convergence. Figure \ref{fig:pressureconvergencecompilation}(a) presents the pressure field of the reference case at $t=6D/V_b$ for the spatial convergence analysis. Let $e_{i}=({p}^{\textrm{ref}}(i)  - p(i)) / V_b$, define the error with ${p}^{\textrm{ref}}$ the interpolated reference data and $p$ the pressure field of a coarser grid. Here, we note that the pressure nodes that reside in the particle's interior are excluded from $e_{i}$ to measure only the convergence from the pressure field surrounding the particle. By doing so, we find the spatial convergence to be of $\mathcal{O}(\sqrt{x})$ for the  $L_2$ norm and zeroth-order for the $L_\infty$ norm. The absence of convergence for $L_\infty$ was also observed by \cite{stein2017} for a cylinder in a prescribed flow. They addressed the lack of smoothness (in the vicinity of the particle) as the main cause for the reduced convergence and proposed a forcing scheme that makes the solution globally smooth. The latter approach is promising, as the grid only needs to be uniform in a small neighbourhood of the boundary, with the same width as the regularised $\delta$-function, but is not implemented here.

Figure \ref{fig:pressureconvergencecompilation} presents the temporal convergence, where we observe $L_2$ and $L_\infty$ to be of first order, owing to the approximation of $A^{-1}=\mathcal{O}(\Delta t)$, similarly as observed for the velocity field.

\subsection{Validations for a freely moving cylinder}\label{sec:validationsmovingcylinder}
In this analysis, we validate the implemented IBPM against the case of a freely rising or settling cylinder. The reference data is taken from \cite{namkoong2008}. In that study, an implicit coupling approach within a finite element framework was used, with a body fitted mesh and adaptive refinement to resolve the cylinder wake. The settling particle has a particle-to-density ratio of $\Gamma=1.01$ and the rising of $\Gamma=0.99$. Our computations are performed on a domain size of $20D\times 60D$. The grid spacing is constant in the vicinity of the cylinder and complies with $D/\Delta x=50$.

\begin{figure}
	\centering
	\includegraphics[width=0.8\linewidth]{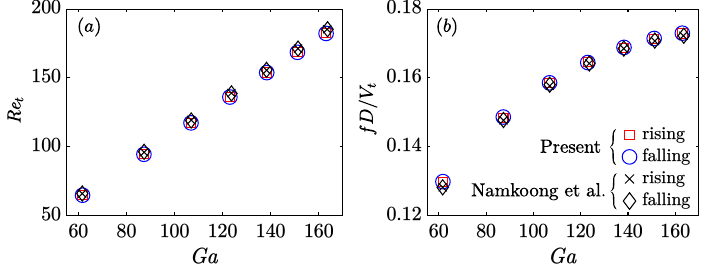}
	\caption{
		Comparison of the present results for a freely falling or rising cylinder with \cite{namkoong2008}.}
	\label{fig:strouhalvsgalileo}
\end{figure}

In figure \ref{fig:strouhalvsgalileo}$(a)$, we present our results of $\Gal$ versus $\Rey_t$, where  $\Rey_t\equiv V_t D/\nu$ is obtained from the terminal vertical velocity $V_t$. Here, we observe a good agreement with \cite{namkoong2008} for the falling and rising cylinders. Next, we compare the corresponding Strouhal number based on the mean vertical velocity of the particle $V_t$ for the same data set. We find the agreement of $\Str$ to be very good as well.

\section{On the value of $I^*_a$}\label{app:theor.anal.istar}
In this appendix we address how the added inertia $I_a^*$ may be estimated, with $I_a^*$ the added inertia due to the surrounding Stokes layer. The torque induced by this layer is estimated for a body that starts spinning in a quiescent viscous fluid such that squares of the velocity field may be neglected. For a sphere it may be shown that this layer yields an effective torque that can be analytically expressed. For a cylinder the analytical solution takes the form of an infinite series expansion \citep{basset1888}. For our analysis we shall assume the torque of the cylinder to take the same form as that of a sphere. The analysis further applies under the condition that the relevant timescale is small enough such that effectively only the history torque contributes \citep[see \eg][]{feuillebois1978,Auguste2018}. Here, we introduce a fitting parameter which we estimate through a series of numerical experiments to match the history torque to that of the cylinder.

\subsection{History torque of a sphere}\label{sec:hist.force.sphere}
 Here, we will present the main results for the history torque of a sphere, which is the dominating contribution of the Stokes layer (see \eg \citet{Auguste2018}), and then find an approximate analogue history torque for a cylinder.
For small time $t$ the leading contribution of the history torque for a sphere is \citep[see e.g.][]{feuillebois1978}
\begin{equation}\label{eq:hist.torque.sphere}
   T_h \approx -\dfrac{1}{6}\sqrt{\pi}\mu D^4 \nu ^{-1/2}\int_0^t\dfrac{d\omega/d\tau}{\sqrt{t-\tau}}d\tau,
\end{equation}
with the rotational velocity $\omega$ assumed to be continuously differentiable. Discretising the integral in \eqref{eq:hist.torque.sphere} for small $t$ one finds
\begin{equation}
    T_h \approx   -\dfrac{1}{3}\rho_f\sqrt{\pi\nu}  D^4  \Delta\omega/\sqrt{\Delta t},
\end{equation}
with $\Delta \omega \equiv \omega^{n+1} - \omega^n$, and the difference between time levels $n$ and $n+1$ being equal to $\Delta t$. If one then introduces
\begin{equation}
\delta = \sqrt{\nu \Delta t / (\pi D^2)}
\end{equation}
one can write for the rotational equation \citep[\cf][]{Auguste2018}
\begin{equation}\label{eq:stokes_layer_rotation_eq}
        \dfrac{1}{10}\Gamma\dfrac{\Delta \omega}{\Delta t} \sim - 2\delta\dfrac{\Delta \omega}{\Delta t}.
\end{equation}

\subsection{History torque of a cylinder}\label{sec:hist.torq.cyl}
Our goal is to find the history torque $T_h$ acting on a cylinder. Here, we assume that $T_h$ approximately takes a similar form as that of a sphere described in \S\,\ref{sec:hist.force.sphere}. To this point we consider a rotating cylinder in a viscous fluid that is at rest initially. The flow is assumed to be axisymmetric and squares of the velocity field are neglected, yielding the expression for the azimuthal velocity component $\ut$  (non-dimensionalised with length scale $D$ and velocity scale $\omega D$)
\begin{equation}\label{eq:ut}
  \Rey_\theta\partial_{\hat{t}} \ut = \partial_{\rh}^2 \ut +\rh^{-1}\partial_{\rh}\ut  - \ut/\rh^2,
\end{equation}
with $\Rey_\theta\equiv 0.25\omega D^2/ \nu$.
Here, we solve \eqref{eq:ut} numerically, by using the fourth order Runge--Kutta scheme for the time-discretisation and a second order central difference scheme for spatial gradients. In addition, we point out  that $\omega$ was given a fixed value when solving \eqref{eq:ut}.
We selected multiple $\Rey_\theta\approx O(1)$ and integrated up to times such that $t\ll D^2/\nu$ (the used time interval ranged  from $10^{-4}D^2/\nu$ down to $10^{-8}D^2\nu$). In this time-interval we assume the history torque
\begin{equation}\label{eq:approxScalingTorque}
    T_h \approx -0.25 \pi D^3 \mu c_1\, \dfrac{\omega}{\sqrt{\pi\nu t}},
\end{equation}
to be dominating the torque on the cylinder. The torque on the cylinder in \eqref{eq:approxScalingTorque} has the fitting parameter $c_1$, which we find by calculating the actual torque on the cylinder from our numerical experiment via $T=0.25D^2 \mu \omega
\int_0^{2\pi}  [\partial_{\rh} \ut - \ut/{\rh}]_{\rh=D/2}\,  d\theta$. Alternatively, the analytical expression of the viscous torque derived by \cite{Mallick1957} may be used:

\begin{equation}\label{eq:actual_torque_cylinder}
    T = 4\pi\mu \left[
    (\omega_0 - \omega) a^2 + \dfrac{\omega a^3}{\pi}
    \int_0^\infty
    \dfrac{y_1(ax)J_0(ax) - y_0(ax)J_1(ax) }{J_1^2(ax) + y_1^2(ax)} \,e^{-x^2\nu t}dx
    \right],
\end{equation}
with $\omega_0$ the initial rotational velocity of the cylinder, $a$ the radius of the cylinder, and $J_n$, $y_n$ Bessel functions of the first and second kind, respectively. Given that the initial field is quiescent, we have the initial cylinder rotational velocity $\omega_0=0$.

In the limit of time $t\rightarrow 0$, we find for the analytical expression and numerical experiment that $T_h = T$ when the fitting parameter takes the value of
\begin{equation}
	c_1 = 1 .
\end{equation}
Now that we have an approximate form of the torque on the cylinder, we can find the analogue form of \eqref{eq:stokes_layer_rotation_eq} for a cylinder. We then plug the obtained expression for the torque in the angular momentum balance and find
\begin{equation}
    \Gamma \dfrac{\pi}{32} D^4 \ddot{\theta} \sim - 0.25 \sqrt{\dfrac{\pi\nu}{t}}\omega D^3.
\end{equation}
To convert the expression for continuously differentiable $\omega$ we apply Duhamel's principle and find (by approximating $\Delta\omega=\omega^{n+1}-\omega^n$)
\begin{equation}\label{eq:rotat.sim.auguste}
    \Gamma \dfrac{1}{8} \dfrac{\Delta \omega}{\Delta t} \sim - 2.00 \delta \dfrac{\Delta \omega}{ \Delta t}.
\end{equation}
The expression in \eqref{eq:rotat.sim.auguste} teaches us that the added inertia $I_a^*$ takes the form of
\begin{equation}
    I_a^* = 16\delta.\label{eq:16delta}
\end{equation}
\subsection{Dataset fit and comparison}
The moment of inertia for an annulus is obtained via
\begin{equation}
    I_a = \dfrac{1}{2} \rho \pi (r_2^4 - r_1^4),
\end{equation}
 with $r_1 = 0.5 D$ and $r_2 = (0.5 + c_1/\sqrt{\Gal})D$, and $c_1$ a constant. Plugging in the latter radii and calculating $I^*_a \equiv I_a/I_{f}$ yields
 \begin{equation}\label{eq:istar}
    I^*_a = \dfrac{8c_1}{\sqrt{\Gal}} + \dfrac{24c_1^2}{\Gal} + \dfrac{32 c_1^3}{\Gal^{3/2}} + \dfrac{16 c_1^4}{\Gal^2}.
\end{equation}
We fitted $c_1$ such that the rotational data depicted in figure \ref{fig:fluidInertiaEffects}({\it a\/}) collapses with respect to $T^*$ (presented in \ref{fig:fluidInertiaEffects}{\it b\/}). For this fit we found $c_1\approx2.3$ to yield good results for cylinders.
To compare with the history torque value from the analysis in \S\,\ref{sec:hist.torq.cyl}, we follow
 \citet{Mathai2018} and set the timescale as half the oscillation time yielding $\Delta t =  0.5 \Str D/V_b$. From this it follows that $\delta = (2\pi \Str\Gal)^{-1/2}$.  By plugging this relationship into \eqref{eq:16delta} and assuming $\Str \approx 0.11$ (the value for resonance where rotation is dominant), we find a value $c_1 = 2.4$, which does match the leading term of our fit ($c_1=2.3$) in \eqref{eq:istar} surprisingly well. In previous work by \citet{Mathai2018} only the leading order term was taken into account, however the inclusion of the additional higher order terms in (\ref{eq:istar}) results in a large mismatch in the actual value of $I^*_a$; more than a factor 2 at a Galileo number of 50. However, we found that inclusion of the higher order terms resulted in a better collapse of the data presented in the current work, their inclusion is therefore recommended.

\section{Comparison to previous work}
\label{app:zhu_mathai}
\begin{figure}
	\centerline{\includegraphics[width=1.0\textwidth]{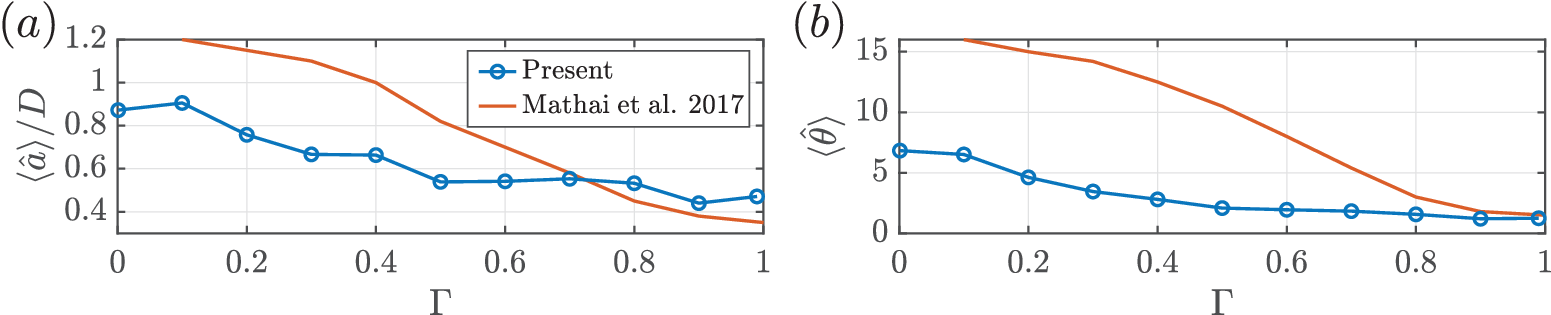}}
	\caption{({\it a\/}) Path and ({\it b\/}) rotational amplitude for Galileo 500 cases without COM offset and $I^* = 1$ compared with results from \citet{mathai2017} (extracted from their figures).}
	\label{fig:Ga500}
\end{figure}
For the cases at $\Gal = 500$ with zero offset, we examined a parameter space spanning $\Gamma$ and $I^*$ that was already explored extensively in the work by \citet{mathai2017}. Here, we take a closer look into the differences between that work and the present one.

We compare our result for $Ga = 500$ with $\gamma = 0$ and $\Gamma$ ranging from 0.001 to 0.99 to results of \citet{mathai2017} at identical parameters. A note on the difference in convention: in the work by \citet{mathai2017} the parameter $I^*$ is equal to $I^*\Gamma$ in the present work. We extracted the results from \citet[figures\,2\,{\it a,\,b\/}]{mathai2017}, where, due to the difference in the definitions of the nondimensional rotational inertia, our results lie on the diagonal $m^* = I^*$, i.e. the line from (A)  to (D). One of the main findings in \citet{mathai2017} was change in the vortex shedding mode from a 2S mode at high $\Gamma$ and $I^*$ to a 2P mode at low values of these two parameters. We found no such transition as all cases in the present work exhibited a 2S mode. Furthermore, \citet{mathai2017} reported that this transition was accompanied by large increases in the path and rotational amplitudes, $\hat{A}/D$ and $\hat{\theta}$, respectively. A direct comparison between the results for these two parameters is presented in figure \ref{fig:Ga500}\,({\it a,\,b\/}). 
While the general trends of a gradual increase for decreasing $\Gamma$ (i.e. low $I^*\Gamma$) for both amplitudes is consistent between the works, there are large deviations in the magnitudes of both translational and rotational amplitudes for identical cases, especially towards lower density ratios.

These differences for identical parameters combinations raise the question if their employed virtual mass approach \citep{schwarz2015} could explain these variations. The use of the virtual mass approach (VMA) was required to stabilise the explicit scheme in \citet{mathai2017}. We tested this hypothesis 
by modifying equations \eqref{eq:KKT_translation}, \eqref{eq:KKT_rotation} to include a virtual mass contribution on both sides scaled by coefficient $C_v$ (for which we discretised the added time derivatives on the right hand side with a forward Euler scheme)
\begin{subequations}
\begin{equation}\label{eq:KKT_trans_VMA}
    (\Gamma + C_v) \dfrac{\textrm{d}\vbg}{\textrm{d}t} =  \dfrac{\Fb_f}{m_f} + (1-\Gamma)g\eb_z + C_v \dfrac{\textrm{d}\vbg}{\textrm{d}t}, 
\end{equation}
\begin{equation} \label{eq:KKT_rot_VMA}
\dfrac{1}{8}(\Gamma+C_v)D^2 I^* \dfrac{\textrm{d}^2 \theta}{\textrm{d} t^2} = \frac{1}{m_f}T_f + \dfrac{1}{8}C_vD^2 I^* \dfrac{\textrm{d}^2 \theta}{\textrm{d} t^2}.
\end{equation}
\end{subequations} 
A value of $C_v = 0 $ corresponds to the present approach, while typical values to stabilise explicit schemes are on the order of the added mass, i.e. $C_v = 1$ for a cylinder \citep{schwarz2015}. We did not observe appreciable changes in the particle dynamics when varying $C_v$ in the range $C_v\in [0,5]$ and it therefore appears that the discrepancies between our work and \citet{mathai2017} are not related to the VMA and may be caused by other unknown factors.

\bibliographystyle{jfm}

\bibliography{Assen2024}

\end{document}